\documentclass[aps,prl,twocolumn,showpacs,superscriptaddress,groupedaddress]{revtex4}
\usepackage{graphicx}  % needed for figures
\usepackage{dcolumn}   % needed for some tables
\usepackage{bm}        % for math
\usepackage{amssymb}   % for math
\usepackage{times}
\usepackage[utf8]{inputenc}
\usepackage[T1]{fontenc}

\newcommand{\be}{\begin{equation}}
\newcommand{\ee}{\end{equation}}

\begin{document}

\title{ Paramagnetic  singularities of the orbital magnetism in graphene with a moiré potential}
\author{J. Vallejo Bustamante}
\affiliation{Universit\'e Paris-Saclay, CNRS, Laboratoire de Physique des Solides, 91405  Orsay, France.}
\author{R. Ribeiro-Palau}
\affiliation{Universit\'e Paris-Saclay, CNRS, C2N, 91120 Palaiseau, France.}
\author{C. Fermon}
\author{M. Pannetier-Lecoeur}
\affiliation{SPEC, CEA, CNRS, Universit\'e Paris-Saclay, 91191 Gif-sur-Yvette, France.}
\author{ K. Watanabe}
\affiliation{
Research Center for Functional Materials, National Institute for Materials Science, 1-1 Namiki, Tsukuba 305-0044, Japan}
\author{T. Tanigushi}
\affiliation{International Center for Materials Nanoarchitectonics, National Institute for Materials Science,  1-1 Namiki, Tsukuba 305-0044, Japan}
\author{R. Deblock}
\author{S. Gu\'eron}
\author{M. Ferrier}
\affiliation{Universit\'e Paris-Saclay, CNRS, Laboratoire de Physique des Solides, 91405  Orsay, France.}
\author{J.N. Fuchs}
\affiliation{Sorbonne Universit\'e, CNRS, Laboratoire de Physique Théorique de la Mati\`ere Condens\'ee, LPTMC, 75005 Paris, France}
\author{G. Montambaux}
\author{F. Pi\'echon}
\author{H. Bouchiat}
\affiliation{Universit\'e Paris-Saclay, CNRS, Laboratoire de Physique des Solides, 91405  Orsay, France.}

\date{\today}

\begin{abstract}
The recent detection  of the singular diamagnetism of  Dirac electrons  in a single graphene  layer %(using a highly sensitive Giant Magnetoresistance sensor (GMR)) %has opened a  sensitive  means  for investigating  
paved a new  way of probing  2D quantum materials through the measurement  of  equilibrium orbital currents  which cannot be accessed  in usual transport experiments. Among the theoretical predictions is an intriguing orbital  paramagnetism at saddle points of the dispersion relation. 
 Here we present  magnetisation  measurements %in a wide range of chemical potential 
in  graphene monolayers aligned on hexagonal boron nitride (hBN)   crystals.  %encapsulated between  two hexagonal boron nitride (hBN) crystals.  One of these is nearly alined with the  graphene  lattice and gives rise to a large period moiré potential.  %acting on the charge carriers of Graphets ne.  
Beside the  sharp diamagnetic McClure response at the Dirac point,   we  detect extra diamagnetic singularities  at the satellite Dirac points  (sDP) of the moiré lattice. Surrounding  these diamagnetic satellite  peaks, we  also observe paramagnetic peaks   located at the  chemical potential  of the saddle points  of the graphene moiré band structure  and relate them to  the presence of van Hove logarithmic singularities in the density of states. These findings reveal the long ago predicted  anomalous  paramagnetic orbital response in 2D systems when the Fermi energy is tuned to the vicinity of saddle points.
\end{abstract}
%\begin{document}
\maketitle

Landau diamagnetism   originates from the   quantum  orbital motion of delocalised electrons at low magnetic field.  
In systems with a  periodic  potential, this orbital magnetism   depends on the  specific  properties of  the lattice.%band structure and the geometrical properties of Bloch wave-functions.
The orbital susceptibility of a single band system is proportional to the  curvature  of the energy dispersion relation (i.e. the inverse effective mass of carriers). This is known as  the Landau-Peierls  result \cite{Landau, Peierls}. In multiband systems, the coupling induced  by the magnetic field between Bloch wavefunctions of different bands   gives rise to new effects. % and  the importance of their topological properties determined by the  fundamental symmetries of the lattice,  was pointed out.
   %A new vision of orbital magnetism  has emerged. %The  geometrical phase encoded in the integral of the Berry curvature over the Brilloin zone and entering in  quantum oscillations of the field dependent magnetisation, also determines the sign of the zero field susceptibility %
The zero field susceptibility is then not  only determined by the curvature of the bands, but also by  geometrical  properties  of Bloch functions such as the Berry curvature  in reciprocal space \cite{Xiao2010, Raoux2014, Piechon2016}.
% (in relation with  the low energy spectrum of Landau levels). Multiband physics plays also  a major role  in this physics : the orbital motion within one band being  influenced by the other  neighboring bands, via  virtual interband transitions mediated by the magnetic field. Such transitions can also be viewed as  geometrical Berry phase contributions \cite{Xiao2005, Xiao2010, Xiao2012} to orbital magnetism.
For instance, a divergent diamagnetism  of  graphene  at the Dirac point (DP), (the touching  point   between electron and hole cone shaped bands) %associated with  A and B crystal sublattices,
  was  predicted by McClure \cite{McClure1956} and linked to the   anomalous $\pi$ Berry  phase   which leads to a zero energy Landau level in magnetic field \cite{Mikitik,Fuchs2010}.% at the crossing point of the 2 bands. %  the square root magnetic field dependence of Landau levels specific of  the conical  dispersion relation.  
%It was also shown that the diverging low field  orbital magnetisation can be instead paramagnetic  when the Berry phase is equal to $2\pi$ with the same conical band structure but no Landau level at zero energy \cite{Raoux2014}. Coming back to graphene,
 %The full chemical potential dependence of the orbital susceptibility has  also been calculated. 
\begin{figure}[!ht]
\centering
\includegraphics[width=\linewidth]{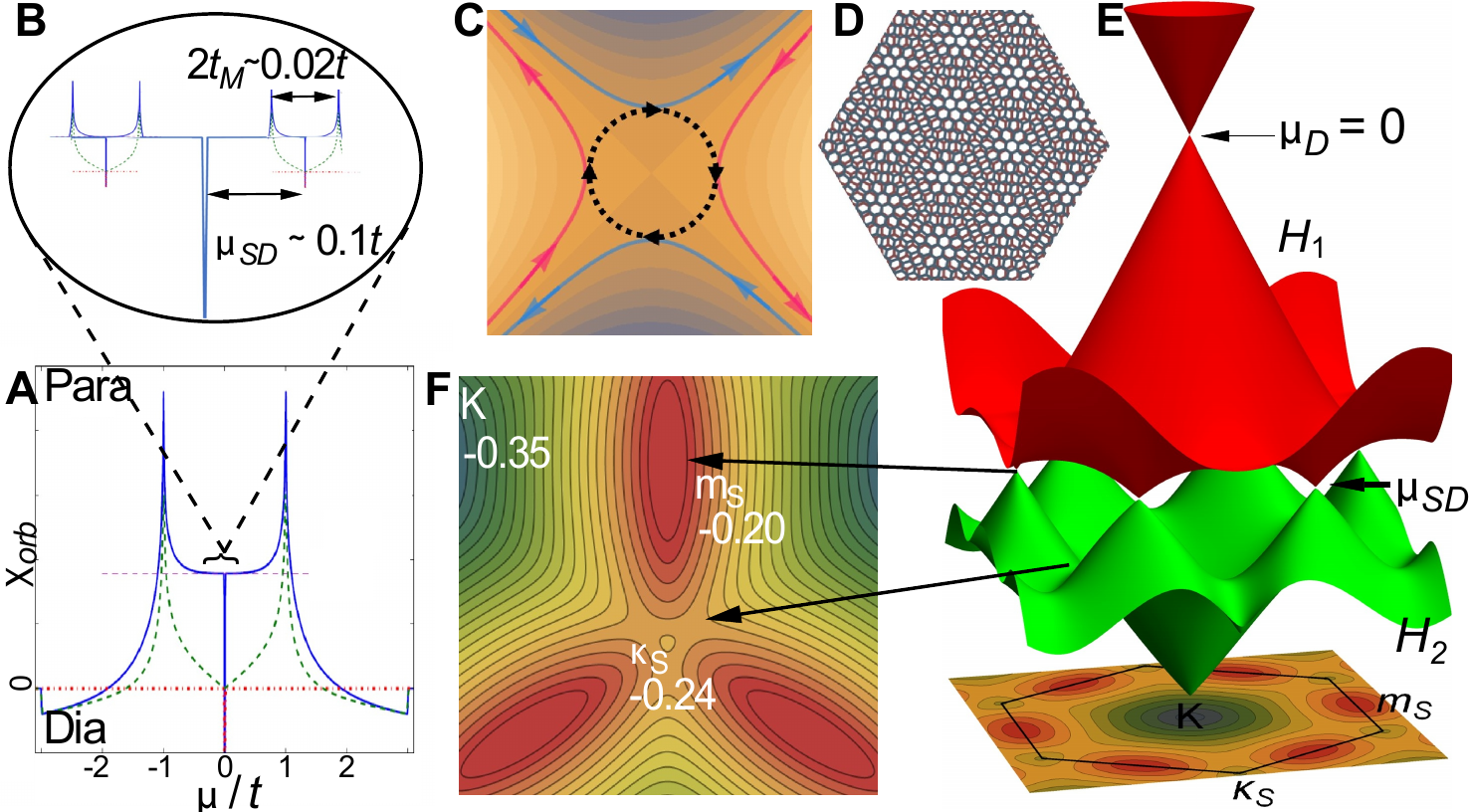}
\caption{ (\textbf{A}) Orbital susceptibility of graphene. Diamagnetic divergent susceptibility is expected at $\mu=0$ and paramagnetic divergences at  the saddle points at  $\mu=\pm t$, (adapted from \cite{Raoux2015}).
(\textbf{B}) Qualitative expectation of the orbital susceptibility of  a graphene/hBN moiré   at low energy  as a function of the chemical potential. (\textbf{C})  Explanation of the existence of paramagnetic currents in the reciprocal space  close to a saddle point in a 2D crystal (adapted from \cite{Vignale91}). (\textbf{D}) Schematic representation of a moiré lattice obtained with the superposition of two honeycomb lattices of different  periods. (\textbf{E}) Mini-band structure  obtained from  the diagonalisation of  the low energy hamiltonian of graphene in the presence of a moiré potential of amplitude $ t_M= -23$ meV. The two highest   energy  hole bands    are represented  below  the main graphene Dirac point. They display  satellite Dirac points at the $m_S$ points of the mini-Brillouin zone. (\textbf{F})  Iso-energy lines of  the lowest hole band ($H_2$), a small dip  surrounded by 3 saddle points is identified at   the $\kappa_S$ point with $C_3$ symmetry. %  Below  $\mu_S$, this $F$ symmetric saddle point split into 3  ordinary $A_1$ saddle points \cite{Fu2020} along the $m_S, \kappa_S$ axis and  an extra satellite Dirac point  at the $\kappa_S$ point appears (see SM).
}
\label{4fig1_OMM}
\end{figure}                                                                                         

It was also predicted that  orbital magnetism can  be paramagnetic rather than diamagnetic. In particular, graphene   is expected  to exhibit two paramagnetic characteristics: i) a  paramagnetic plateau \cite{Stauber2011,Raoux2014, Raoux2015, Piechon2016} on either side of the Dirac point; and %when doping is increased both on the electron and hole sides with, in addition,and 
ii) logarithmic paramagnetic divergences   
 when the Fermi energy coincides with  saddle points of the graphene band structure  \cite{Raoux2014}, see Fig.1A.  Such  paramagnetic orbital susceptibility    peaks proportional to the van Hove (vH) singularities in the density of states (DOS) %as a function of the  chemical potential $\mu_F$ 
 $\rho(\epsilon)$, were predicted   long ago at saddle points of the band structure  of any 2D   crystalline materials  by Vignale \cite{Vignale91}. %(We outline in SM the similarities between the calculation of the susceptibility and the DOS which involve both  identical energy integrals  explaining the similarity between their singularities.)
 %This author provided also  
A simple physical explanation of  the  paramagnetic sign lies in  %the coexistence of 
 the opposite signs  of effective masses  $m_x$ and $m_y$   at the saddle points and the fact that  the Landau-Peierls susceptibility is proportional to $-\rho(\epsilon)/ m_x m_y$. In a magnetic field, near a saddle point, carriers follows hyperbolic-like trajectories in reciprocal space.
 Tunneling  between these trajectories (also called magnetic breakdown \cite{Falicov}) gives rise to reconstructed quasi-circular paramagnetic trajectories around the saddle points, see Fig.1C. Reaching these saddle points in pure graphene requires  doping to unattainable Fermi levels  of the order of the nearest neighbor hopping energy  $t=2.7$~eV.  %$t=2.7$~eV, by aligning cristallographically graphene with hexagonal boron nitride (hBN) it is possible to induce a large periodicity  potential. This will modify the electronic band structure generating, among others, saddle points at reachable energies%. 
However, in the following we show  how, by inducing a large wavelength  moiré periodicity  in graphene   aligned to a hexagonal boron nitride  crystal (hBN), we can reach such saddle points in the  moiré band structure  at reasonable doping 
and  detect the expected  singular paramagnetic orbital response. %Depositing  graphene on hexagonal boron nitride gives rise to a  6-fold symmetric potential  acting on charge carriers.% This quasi periodic potential is characterized by  a  lattice parameter 

     Due to the small difference between hBN and graphene lattice parameters, the moiré lattice parameter $a_M$ of graphene aligned  on hBN, is   much larger than the size of the unit cell of graphene (see Fig. \ref{4fig1_OMM}D). % $a_G =0.25$ nm, reaching  14 nm  when the   lattices are perfectly  aligned. 
The  large period moiré potential  leads to the formation of low energy minibands %with a reduced Brillouin zone  whose size is of the order of    $ k_M= |\vec {K \kappa_S}|= 2\pi a_M ^{-1}$ 
 centered on each Dirac point  (see Fig. \ref{4fig1_OMM}E), and the occurrence of low energy satellite Dirac points   (at $\mu_{SD} \simeq \pm t/10$). These sDPs ,  accessible by applying  moderate gate voltages, were observed experimentally by several groups including \cite{Yankowitz, Dean2013, Woods2014, Hunt, Ribeiro}. % The twist angle dependence of satellite Dirac points was evidenced  via  transport experiments in gated  hBN/Graphene samples  \cite{Ribeiro} and 
They are surrounded  by saddle points whose associated vH singularities   were detected via  DOS measurements \cite{Yankowitz}. Saddle points were also revealed in electron focusing experiments \cite{Goldhaber2016} and more indirectly in  magnetic   field dependent patterns  in  Josephson junctions \cite{Delagrange}. %revealed   on both sides of sattelite peaks. 
Field dependent  peaks  in  photo-emission  spectra \cite{Wu2016}  as well as in  thermoelectric Hall measurements \cite{Moriya} %(measured as a function of the gate voltage) {golhaber}
were interpreted as related to   orbital magnetisation  peaks at those low energy vH singularities. 

 In this letter we present  direct magnetisation measurements on graphene/hBN moiré  samples in a wide range of chemical potential. Our experiments reveal the   paramagnetic susceptibility singularities  predicted  long ago   at saddle points  of the moiré dispersion relation. %on both sides  of diamagnetic peaks at the satellite Dirac points independently identified trough transport measurements.

  %(The situation where the  satellite Dirac points are located on  the $M$ points of the reciprocal superlattice, is specially investigated in this work. It   gives rise to   $C_3$ symmetric saddle points   with van-Hove singularities diverging in $\epsilon ^{-1/3}$ . This is  in contrast  with the ordinary $A_1$ symmetric saddle points  with logarithmic divergences of the density of states.) %expected instead when he moiré potential is treated at a perturbative level and satellite Dirac points are located  at the $\mu$ points of the reciprocal moiré lattice.

%As in our recent experiments  which could  reveal the singularity of the Landau magnetism of graphene in the vicinity of the Dirac point \cite{Vallejo2021}, we use specially designed   pairs of giant   magnetoresistive sensors which detect the stray field of the  magnetic moment of  gated graphene samples  deposited between the 2 sensors. The  sensitivity of the set-up is of the order of 0.1nT at perpendicular field of 0.1T. These sensors are connected in a Wheatstone bridge configuration in order  to cancel all the magnetic contribution from the environment. The top gate is modulated at a frequency of 120Hz to enhance  the sensitivity  of the detection and eliminate spurious magnetic contributions independent of the chemical potential of graphene. The signal obtained is the derivative  of the magnetisation of graphene  with respect to $V_g$.

{ \it Moiré samples and magnetic detection}  \newline
\begin{figure}[!ht]
\centering
\includegraphics[width=1\linewidth]{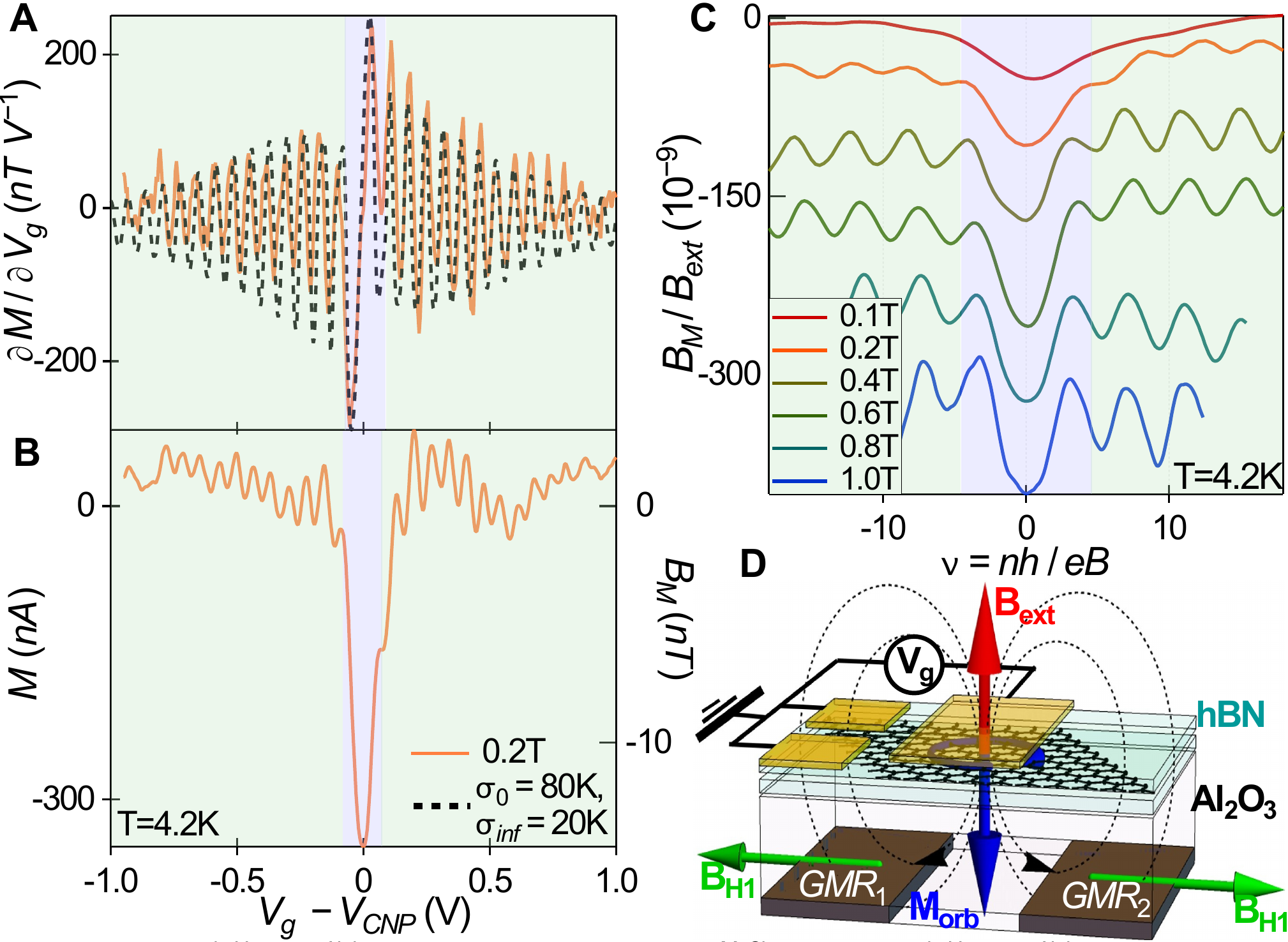}
\caption{ Low doping data on sample $M_A$. \textbf{(A)} Solid line: derivative of the  magnetic field  detected by the GMR detector as a function of the gate voltage $V_g$ close to the DP at $B=0.2$~T. The  curve is the average of 10  independent measurements. Dashed line: theoretical gate dependence of ${\partial M}/{\partial V_g}$. The  disorder amplitudes   $\sigma_0$  and  $\sigma_{\infty}$ were determined  from the width of the  McClure Dirac  and the decay of dHVA  oscillations. They are respectively 80 K and 20 K . \textbf{(B)}   Magnetization per unit surface obtained by integration of the data in (A) with, on  right axis,  units of the equivalent measured magnetic field by the GMR detector. \textbf{(C)} Evolution of the McClure peak  and dHvA oscillations for different applied  out-of-plane magnetic fields. On the y axis the detected magnetic signal   on the GMR detector is normalized by the applied  field. On the x axis $\nu= n \Phi_0 /B$ is the Landau levels filling factor  
where $n$ is the carrier density.% $n =C_g V_g/e $%is obtained from the gate voltage  normalized by the square of the Landau levels energy $\nu =V_g/\alpha\epsilon_B^2$, where $\alpha={e}/{\pi C_g \hbar^2 v_F^2 }$ where $C_g$ is the capacitance per unit surface  of  the gate. 
(The  apparent   increase  of $\chi $ with $B$  at low field is an artefact due to the gate voltage modulation which broadens the McClure peak. This  effect becomes  negligible when the   intrinsic McClure peak  broadening, proportional to B, exceeds the gate voltage modulation.)
\textbf{(D)} Sketch  of the experimental setup. The orbital magnetisation is detected by   the voltage between the two  dc current biased  GMR detectors $R_1$ and $R_2$ %measuring the horizontal component of the field generated by current  orbital loops through the sample. The derivative with respect to gate voltage of the orbital magnetisation of the graphene area below the gate electrode   is  proportional to  the ac component of the voltage difference  
measured by a lock-in amplifier  at the modulation frequency of the gate voltage. }
\label{4fig2_OMM}
\end{figure}
In order  to reach the satellite Dirac points and the neighboring saddle points  at moderate   doping, we fabricated samples where the hBN and graphene lattices are nearly aligned, leading to the maximum value  of the G/hBN moiré super-lattice  parameter as described in SM.  We investigated  two different samples, $M_A$ and $M_B$. Raman spectroscopy was used to verify the alignment (see SM and \cite{Ribeiro, Eckermann})  and determine the  lattice parameters  yielding  $a_{M_A}=9.5$ $\pm0.5$ nm and $a_{M_B}=12.5\pm0.5$ nm corresponding to  1$^{\circ}$ and 0.6$^{\circ}$ mismatch angles respectively. %According to \cite{Ribeiro} (see also \cite{Eckermann}) ,   the FWHM of the 2D peak of the Raman spectrum of graph ene  varies linearly with the size of the moiré superlattice, $a_M$ leading to the relation FWHM$_{2D}=2.7a_M+0.77$. In the case of the samples investigated here, Raman measurements revealed FWHM$_{2DM_A}=26.5\ cm^{-1}$ and FWHM$_{2DM_B}=34.9\ cm^{-1}$ as shown in the SM. This corresponds to $a_{M_A}=9.5nm\pm0.5nm$ and $a_{M_B}=12.5nm\pm0.5nm$.

The magnetisation experiments  were performed using the  technique described in \cite{Vallejo2021}. The encapsulated samples are  deposited on a magnetic field detector based on a pair of highly sensitive giant magnetoresistance (GMR) probes \cite{GMR1,GMR2,GMR3}. The key point of using these  sensors is that whereas the orbital magnetism in the system  is generated in response to the out-of-plane external field, the GMRs are only  sensitive to the in-plane components of the stray field created by the orbital currents. Connecting the two GMR strips in a Wheatstone bridge configuration and modulating the  DC  gate voltage, $V_g$, with an AC bias (Fig.2C), %at a frequency chosen to minimize electronic noise
eliminates most of the extraneous  spurious  magnetic contributions. 
%In addition, a metallic gate is placed on top of the encapsulated graphene  between the GMR probes and swept to precisely control the chemical potential. 
%Moreover the gate voltage  is modulated at a frequency of 120Hz to enhance the sensitivity of the detection and to obtain a signal corresponding to  This allows us to eliminate spurious magnetic contributions independent of the gate voltage. Therefore, the magnetic signal comes from the part of graphene  between the GMRs and covered by the gate.The gate modulation gives access  to 
  In this way, one obtains the $V_g$ derivative of the magnetisation of  graphene area below  the gate  between the GMR probes. Additional electrodes   outside the GMR detection  region allow  for transport measurements and an independent  determination of the   main and satellite Dirac points positions. For an applied out-of-plane magnetic field  of 0.1 T, the in-plane sensitivity of GMRs sensors coupled to $M_A$ and $M_B$ is respectively 2.5 and 1 $\Omega$/mT and their field equivalent noise is about 1nT/$\sqrt{Hz}$. 
%It is of about $6\times 6 \mu m$ and $15\times 6 \mu m$ for samples $M_0$ and M1 respectively. However, as the contacts are outside the GMR region, the transport mesurements may measure a larger zone.
%\begin{figure}[!ht]
%\centering
%\includegraphics[width=0.5\linewidth]{yao.png}
%\caption{First sample. Made at C2N with RRP.}
%\label{sample}
%\end{figure}

% In setting up this template for *Science* papers, we've used both
% the \section* command and the \paragraph* command for topical
% divisions.  Which you use will of course depend on the type of paper
% you're writing.  Review Articles tend to have displayed headings, for
% which \section* is more appropriate; Research Articles, when they have
% formal topical divisions at all, tend to signal them with bold text
% that runs into the paragraph, for which \paragraph* is the right
% choice.  Either way, use the asterisk (*) modifier, as shown, to
% suppress numbering.

{ \it Main Dirac point region: McClure peak and de Haas-van Alphen oscillations.}

%Figure \ref{4fig2_OMM}A shows the derivative of the magnetisation as a function of the gate voltage for the M1 sample, in a perpendicular magnetic field of 0.2T. The sharp anti-symmetric peak at around -0.55V can be identified as the singular diamagnetic response of graphene and next to it, we can identify the region of de Haas - van Alphen (dHvA) oscillations.  The results obtained for the low energy region are in agreement with what was previously reported \cite{Vallejo2021}. From the oscillations region, it is possible to obtain experimental parameters such as the long range disorder scale energy or the thickness of the top BN. The former will allow us to adjust the theoretical model to our sample and the latter lets us link the applied gate voltage to the doping, and thus, confirm the moiré angle through the position of the secondary peaks.
 As shown in  Fig.\ref{4fig2_OMM} the magnetisation close to Dirac or charge neutrality point (CNP) features the diamagnetic McClure peak  \cite{Vallejo2021}. This %gate voltage dependent
 peak  broadens with out-of-plane magnetic field field and  de Haas-van Alphen (dHvA) oscillations  appear  with increasing doping. Fig.\ref{4fig2_OMM} shows both  the derivative of the magnetisation and the integrated  curve as a function of the gate voltage for  sample $M_A$ , in a out-of-plane magnetic field of 0.2 T.  The amplitude of the detected signal  $B_M$  is  15 nT at the DP. This data can be precisely described  deriving the magnetisation from the  field dependence of the grand potential
of graphene at a chemical potential fixed by the gate voltage \cite{ Vallejo2021}.
%and describes  the thermodynamics  of fermions   occupying  the Dirac Landau spectrum, 
%The  Dirac Landau spectrum    is composed of 
The Landau   energy levels in graphene are $\epsilon_N=\pm\sqrt{N}\epsilon_B$ with $\epsilon_B=\sqrt{2ev^2\hbar B}$,  where $v$ is the Fermi velocity and $N$ is an integer \cite{McClure1956}.  Disorder is modeled  by a gaussian distribution of chemical potential $\mu$  whose standard  deviation $\sigma _\mu$ decreases at high doping, due to screening effects which are more efficient. The magnetisation was shown to be a universal function of the variables $\mu /\epsilon_B$ and  $\sigma/\epsilon_B$ .
The  McClure peak at low field has a width $\sigma_0=80\pm 5$ K  at the  CNP.  dHvA oscillations observed at larger doping exhibit  a characteristic energy scale of $\sigma_{\infty}=20$ K. Both $\sigma_0$ and  $\sigma_{\infty}$ are twice smaller than in our previous work \cite{Vallejo2021} and indicate a  better  quality of the present samples.  The dashed curve in Fig.\ref{4fig2_OMM}A is the   theoretical fit for $\partial M/\partial V_g$ using those parameters. %This is why the data presented in this paper is of much higher quality than the one presented in \cite{Vallejo2021}. %Figure \ref{4fig2_OMM}C shows the gate voltage dependence of the magnetisation obtained by integration of data similar to Fig  \ref{4fig2_OMM2}A measured for different values of magnetic field. for the same field, from where the estimated value of magnetisation at charge neutrality point (CNP) was obtained. 
Fig.\ref{4fig2_OMM}C shows the evolution of the  magnetisation  with Landau level filling factor for different field values. The magnetisation  is  renormalized by the  applied magnetic field, which   in the linear regime is the  magnetic susceptibility $\chi=M/B$. One notes the increased  dHvA oscillations relative to the McClure response as  magnetic field increases. % induced by  the gate voltage  modulation. % since the variation of amplitude is most notorious at low field, where the McClure peak and the dHvA oscillations are narrower. 
 
\begin{figure}[!ht]
\centering
\includegraphics[width=1\linewidth]{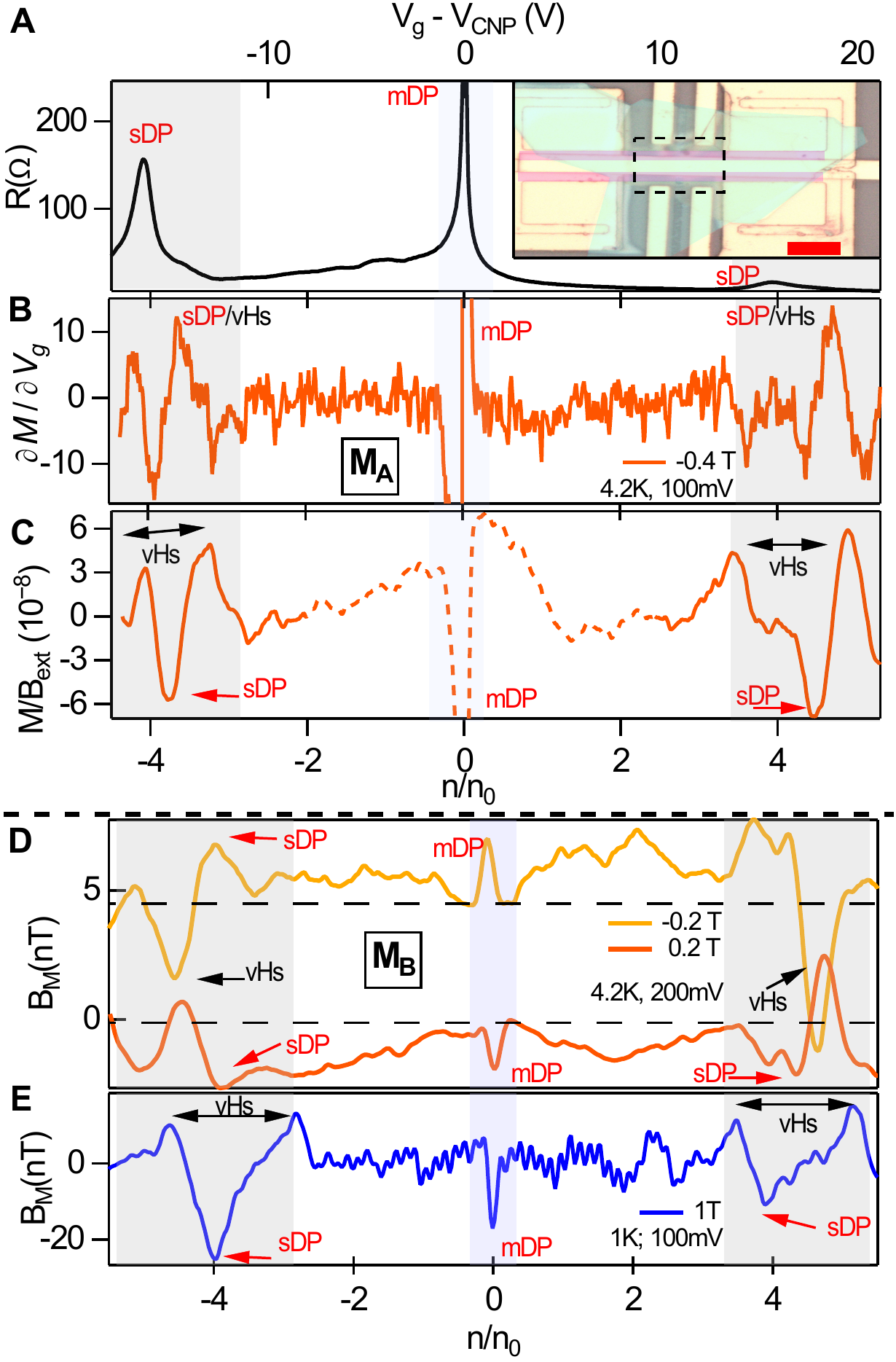}

\caption{\textbf{(A)} 4 terminal resistance of   sample $M_A$.  Inset: optical image of the  sample  $M_A$ on the top of the GMR detector  (red  scale bar is 10  $\mu$ m). \textbf{(B)} Derivative of the magnetisation as a function of the  gate voltage  and carrier density (renormalised to the moiré filling factor $n_0$) for a wide doping range in an external magnetic field of 0.2 T for the same sample. \textbf{(C)} Magnetization (in units of the  magnetic field detected on the GMRs,  renormalised by the out-of-plane applied field)  obtained by  integration of the data in (B). In the region of the secondary Dirac peaks we observe diamagnetic peaks (red arrows) surrounded by paramagnetic peaks (black arrows). \textbf{(D,E)} GMR data measured on  sample $M_B$.  Numerically integrated magnetisation as a function of the carrier density (renormalised to $n_0$ ) at  $\pm$0.2 T,\textbf{(D)} and 1 T, \textbf{(E)}.}
\label{4fig3_OMM}
\end{figure} 

{\it Satellite Dirac points, diamagnetic and paramagnetic singularities.} \newline
We now turn to the higher doping regime. Fig.\ref{4fig3_OMM}A shows the four-terminal resistance of sample $M_A$ in a wide range of gate voltage at 4 K. %We can identify %the main Dirac peak at $V_g=-0.55V$ as well as 
Both satellite peaks are clearly visible at -16 V and 15.5 V from the CNP. %The satellite Dirac peaks at $V_g -V_{CNP}=-16$ V and $V_g - V_{CNP}= +15.5 $ V, symmetrically positioned  with respect to the main Dirac peak  at $V_{CNP}$ are clearly visible. %From these values one can estimate a moiré period $a_M= 9\pm 0.5 nm $ for this sample. 
Fig.\ref{4fig3_OMM}B,  shows the  magnetisation  response at 0.2 T in the same range of gate voltage, using a 100 mV modulation of the gate voltage. %measured on the same sample. 
 This strong  $V_g$ modulation  increases  the detection sensitivity at high gate voltage but, because the chemical potential scales as $\sqrt{|V_g|}$,  damps the previously discussed  diamagnetic McClure response    and  dHvA oscillations. In the high doping region of interest here, in particular in the region where the sDPs are found in the resistance measurement, we find a series of three antisymmetric peaks, compatible with the expected  moiré band  orbital magnetism  as shown below.
%: secondary Dirac peaks should be related with diamagnetic peaks, whereas van Hove singularities should give rise to paramagnetic peaks.
%Figure \ref{4fig3_OMM}B by itself does not allow for a good determination of the signs of the derivatives in the secondary Dirac region. 
The integrated trace displayed  in Fig. \ref{4fig3_OMM}C  features   a diamagnetic peak (red arrow) in the hole doped region  at $V_g - V_{CNP}=-17$ V  surrounded by two paramagnetic peaks (black arrows). In the electron doped region, we find similar features %detect another  diamagnetic peak at the sDP and    one  clear paramagnetic peak at large doping, the low doping side 
showing  though somehow different positions of the peaks with an asymmetry in the position of the two  paramagnetic peaks with respect to the diamagnetic one. From the value of moiré lattice parameter  we find that  the gate voltage  positions of the diamagnetic peaks correspond, as expected, to a carrier density of 4 $n_0$  where $n_0$ is the number of carriers per moiré cell, (the factor 4 comes from spin and valley degeneracies).  The peak positions differ slightly   from  those observed on the resistance measurements shown in Fig.\ref{4fig3_OMM}A, a discrepancy we attribute to the  different  sample region probed in the resistance measurements  \cite {notedetaVg}.  We assign  the paramagnetic peaks to  the  expected  magnetic orbital response at the saddle points of the moiré miniband structure. {\it This is the main result of our letter}. Using the gate capacitance of sample $M_A$,  we determine the energy splitting between the paramagnetic and diamagnetic peaks and therefore the  expected positions of the vHs, to be in the range of 10 to 20 meV. This  yields  an estimate of the  amplitude $t_M$ of the moiré potential (see Fig.1B), as will be discussed  more precisely below.
%To physically understand the origin of this paramagnetic response, it was shown that the interband contribution to the orbital susceptibility follows the logarithmic divergence of the density of states as the Fermi energy approaches the saddle point in a 2D lattice \cite{Vignale91}. This phenomenon, that was first computed for a non interacting 2D square lattice, is found in many 2D crystals, as pristine graphene for example as shown in figure \ref{4fig1_OMM}B. However, the high energy value at which the van Hove singularities are located makes it impossible to attain this doping electrostatically and so, impossible to measure. The long range superpotential of moiré affects the bandstructure of graphene, creating satellite Dirac points that are surrounded by van Hove singularities, whose orbital paramagnetic signature is shown in figures \ref{4fig3_OMM}C (and later again in \ref{4fig4_OMM} for sample $M_0$).

Fig.3D and 3E also presents  equivalent  data on sample $M_B$. %Figure \ref{4fig4_OMM}A shows the 2-wire resistance as a function of gate voltage for the explored region. Due to problems with contacts, the curve does not follow a clean, smooth line as in sample $M_1$, but nevertheless, the main and satellite Dirac peaks are observable. 
%for an external magnetic field of 1T at 1K the diamagnetic McClure peak at the Dirac point and  dHvA oscillations at larger doping. 
   At high hole and electron doping one    identifies  several peaks  of similar amplitude. The position of the  diamagnetic satellite peaks are consistent with transport data (shown in SM) and with a moiré period %of $a_M= 12.5\pm 0.5 nm $ 
 larger than that of sample $M_A$. There as well, HvA oscillations are attenuated  by the gate voltage  modulation of 100 $mV$ and invisible at $\pm 0.2$ T, but   are  visible  at 1 T due to their larger period. %The diamagnetic  peaks at the sDP are surrounded by   paramagnetic singularities expected from  vHs  at the saddle points of the band structure.   %on the same sample and shown in Fig. \ref{4fig1_OMM}B  and C.
The data taken, %one observes diamagnetic satellite Dirac peaks and   the outer paramagnetic peak. 
at $\pm 0.2$ T  display  peaks  of opposite sign,  approximately at the same positions  and  with 5 times smaller amplitude  than the 1 T data  which is consistent with a linear field dependent magnetisation. However, whereas one observes clearly  diamagnetic satellite Dirac peaks and   outer paramagnetic peaks, in contrast with the  data on $M_A$  the inner  smaller paramagnetic peaks are nearly undetectable at $\pm 0.2$ T whereas they are visible at 1T. We note that  the magnetic energy scale   $\epsilon_B$ is equal to 30 meV at 1 T i.e. of the order of the moiré potential, which means that the miniband spectrum  is modified in a non perturbative way at this field. This can explain why the data  at 1 T is significantly different from the    %show the gate voltage dependence of the magnetisation
the  lower field data. 
In the following, we   discuss   explanations for  the  asymmetry  in amplitude of paramagnetic singularities  on  either side of satellite Dirac peaks. 

%The data taken, %one observes diamagnetic satellite Dirac peaks and   the outer paramagnetic peak. 
 %at $\pm 0.2$ T  display  peaks  of opposite sign,  approximately at the same positions  and  with 5 times smaller amplitude  than the 1 T data  which is consistent with a linear field dependent magnetisation. However, in contrast with the 1 T data,  the inner   paramagnetic peaks are nearly undetectable at $\pm 0.2$ T.

 %  Asymmetries   were also observed in vH singularities detected through DOS  experiments \cite{Yankowitz}.    

\begin{figure}[!ht]
\includegraphics[width=1.\linewidth]{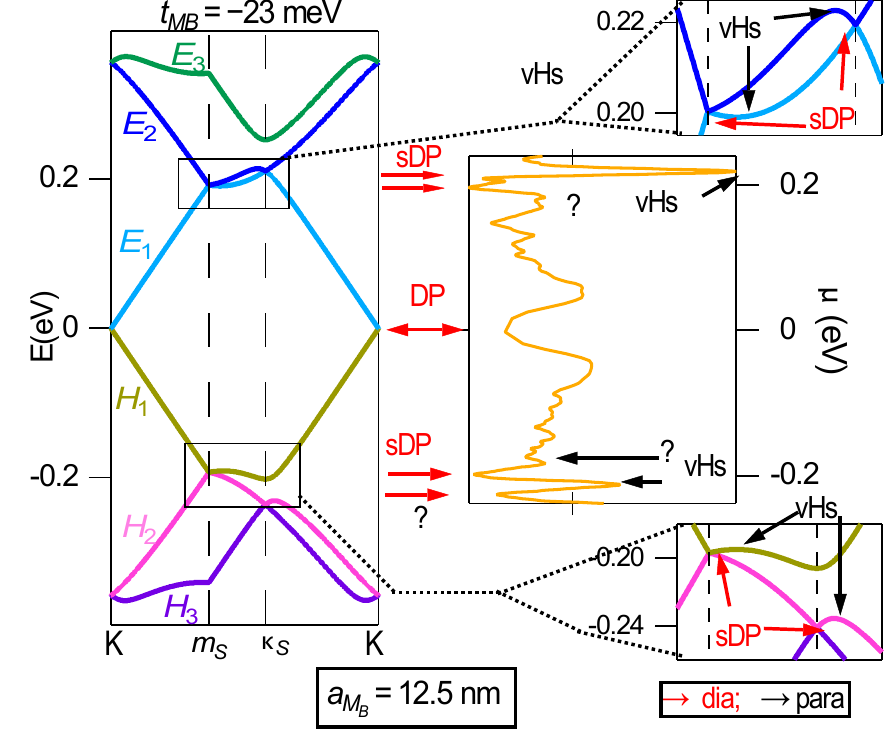}
\caption{   %(We  note due to  the  $\sqrt{V_g}$ dependence, the energy width of satellite Dirac points is  smaller than the width of the main Dirac point.) 
Comparison between band structure (left) and experimental data for sample $M_B$ (right).  We present cuts along the  three $K,m_S$,  $m_S,\kappa_S$, and  $\kappa_S,K $  axes of the moiré band structure calculated for  $t_M =-23$ $meV$  matching the position of the observed diamagnetic peaks (red arrows) and paramagnetic peaks  (black arrows) in  magnetisation data   function of the  chemical potential. 
\label{4fig5_OMM}}
\end{figure}

{ \it Comparison with a simple theoretical model}\newline
Computations of moiré spectra rely on  specific modelisations of the moiré potential \cite{Yankowitz,Falko,MacDonald,Koshino}. %Initial perturbative calculations from the hopping integrals between hBN and C atoms \cite{Yankowitz} demonstrated  the occurrence of satellite Dirac points. More subtle effects due to  lattice distortions induced by the interaction between the hBN  and graphene layers can also be included.  They open a small gap  at the  main and satellite Dirac points \cite{Falko, MacDonald}. Experiments  show that these distortion effects  are more important for very small angles between the hBN and the graphene layers and and quite weak for larger twist angles for which the perturbative model of \cite{Yankowitz} describes  satisfactorily the physics. %we estimate the positions and amplitudes of the  different susceptibility peaks  %experimentally observed in the vicinity of the secondary satellite peaks using
We  use the simplest model which reproduces  the positions and amplitudes of the susceptibility peaks at the Dirac and saddle points.%the primary focus of our experimental measurements.
This model, initially derived in  \cite{Yankowitz}, assumes a $C_6$-symmetric moiré potential of amplitude $t_M$ which only depends   on the  positions of carbon atoms with respect to the hBN atoms. We neglect all  contributions  breaking the inversion symmetry of the graphene lattice  such as considered in \cite{Falko, Wu2016}. The miniband spectrum,  folded into the  moiré Brillouin zone,  can be easily calculated within  this approximation as shown in Fig.1, Fig.4 and SM. The amplitude of the moiré potential determines the number of satellite peaks and their position in the momentum space as well as the symmetry of the saddle points. These features also depend on   the minibands considered  and the type of carriers, electrons or holes, see \cite{ Yankowitz,Falko, Wu2016} and theoretical discussion in the SM.  When increasing the  amplitude $t_M$ of the moiré potential, one finds a clear  electron-hole asymmetry of the   minibands spectrum   determined by the sign of $t_M$.   This is illustrated in Fig.4  showing   cuts along the   $ K, \kappa_S,m_S, K$ axis of  the three  lowest  moiré electron and   highest hole bands   around  the   main  DP. For negative values of $t_M$,   the moiré minibands are  found to be wider on the hole side compared to the electron side.   Whereas crossings occur essentially  between the first and the second bands on the electron side, they also occur between the second and third bands on the hole side. These crossings determine  the  number and  position of the satellite Dirac points in  reciprocal space.  % they stand  either  at $m_S$ or at $\kappa_S$ points of the  moiré Brillouin zone  depending  both on the amplitude and the sign of $t_m$. 
When $|t_M|$  is smaller than $25$ meV, band crossings occur  both at  $\kappa_S$ and $m_S$ points for the first two electron  bands (see SM ). % the low energy  crossings occur only at $m_S$ points on the hole side.
These sDPs are separated by ordinary  named $A_1$ saddle points with a $C_2$ symmetry  that connect 2 valleys  \cite{Fu2020}. %We show in SM that the  logarithmic vH singularities are modified when the  absolute values of the curvatures at the saddle points differ and exhibit a crossover from a logarithmic divergence to a $1/\sqrt{\delta\mu}$   for positive $\delta \mu = \mu -\mu_S $  where $\mu_S $ is the  chemical potential  at the saddle point.

  Coming back to experiments,  we determined $t_M$   from the position of the measured diamagnetic and paramagnetic satellite peaks  compared to  their expected positions  	in  the  moiré band structure  for both  samples. %(Good agreement is obtained for   values of  $t_m $ which differ  in amplitude on  the hole and  electron sides.  We attribute this extra electron-hole asymmetry  to  charged dopants, a known explanation  for the  observed asymmetry of the main Dirac  McClure peak of graphene. (Electrostriction effects in hBN \cite{piezo} could also be a cause). 	
 The calculated spectra  which best match  the experimental  positions of the different peaks  are obtained for $t_M=-15$ meV and  $t_M=-23$ meV for samples $M_A$ and $M_B$ as shown in SM and Fig.4 for sample $M_B$. Because  of these relatively small values of $t_M$,  $E_1$ and $E_2$ bands cross  both at $\kappa_S$ and $m_S$ points  at different energies, we  therefore expect  two families of sDPs.  This is compatible with the  split diamagnetic satellite  McClure peak shown   in Fig.4 on the electron side.  We attribute the larger splittings observed on the hole side to the crossings between  $H_2$ and $H_3$ bands.

A more careful analysis of  the shape  and curvature of the minibands  Fig.4,  %together with the analytical derivation of their  dependence
 in the vicinity of the $\kappa_s$  and $m_S$  satellite Dirac points  (see also SM) helps explain the asymmetry in  position and amplitude  of the paramagnetic singularities. These are  clearly  more pronounced on the high doping sides of sDPs. We find that the miniband curvatures at the $A_1$ saddle points along  $[ m_S, \kappa_S ]$ are  much larger  in bands  bands $E_2$ and $H_2$ compared to bands bands $E_1$ and $H_1$. As a result the position in energy of the saddle points on the low doping minibands ($E_1$ and $H_1$) are very close to the $m_S$ sDP , in contrast with  the large doping ones ($E_2$ and $H_2$) located further in energy above $\kappa_S$ .   We show in SM I.C and E  that  these observations  explain why the outer paramagnetic peaks    are more intense than the inner ones (tending to overlap with the diamagnetic ones at the $m_S$ SDPs),  both on  electron and  hole side. This asymmetry is more pronounced for sample $M_B$ whose moiré potential is  larger than for sample $M_A$. 
 
We also find  theoretically that the first   hole miniband $ H_1$ below the main  Dirac peak  does not exhibit a Dirac point at $\kappa_S$ but  instead  a  broad shallow  minimum turning into a plateau for $|t_M| \geq 50$ meV  as shown in SM Fig.S3  and can be viewed   as the merging of three $A_1$  saddle points.  This is the signature of a $C_3$ saddle point  surrounded by 3 maxima separated by 3 valleys at 120 degrees and  characterized by a dispersion relation varying as a cubic function of the wave-vector  (2nd order curvature cancels in all directions, see  \cite{Fu2020}, for the classification of saddle points). In SM I.D we show that  at  $C_3$ saddle points,  the  DOS  exhibits a  $\delta \mu ^{-1/3 }$  divergence  in contrast with the logarithmic divergence at $A_1$ saddle points in agreement with \cite{C3}. Moreover, since the effective mass diverges, the Landau-Peierls susceptibility is then  not simply proportional to the density of states as shown in \cite {Vignale91}. The calculation  of the susceptibility is  more involved,   leading to a $-\delta\mu ^{1/3 }$ singularity in contrast with the logarithmic divergence   for $A_1$ saddle points \cite {Vignale91}.

%When the amplitude of $t_m$ si large enough  	minibands respectively above  the hole and below the electron SD Dirac points exhibit $C_3$  saddle points, %whereas   we  note  for the minibands below  SD points, the presence of  sharp extrema  at K points either a maximum or a minimum  which correspond  to extra Dirac points at the intersection between the third and second lowest minibands. %attention pas toujours vrai (CF Wu et al.) et cela dépend de l'amplitude de $t_m$).
   
%Their positions in energy are different on the  electron   compared to the  hole doping side as 
 %Different  shape of saddle points  are also observed in the miniband spectrum. The  $C_3$ symmetric saddle points centered  at   $\kappa_S$ points  are destroyed  for minibands for which an extra  gaped Dirac points at K points shows up, leading to a splitting   of these $C_3$ saddle points into three ordinary  saddle points between $m_S$ and $\kappa_S$ points as already noted in \cite{Wu2016}.
 %The situation where the  satellite Dirac points are located on  the $M$ points of the reciprocal superlattice, is specially investigated in this work. It   gives rise to   $C_3$ symmetric van-Hove singularities  with a power law divergence in $\epsilon ^{-1/3}$. This is  in contrast  with the ordinary $A_1$ symmetric van-Hove singularities with logarithmic divergences. 

At present our  experiments are not accurate  enough   to  identify precisely the nature of the different saddle points,  but we can  compare   the   amplitudes of the measured magnetisation singularities to  theoretical predictions.   %We restrict this comparison to sample $M_B$ which has the largest   $a_M$ and $|t_M|$. 
We first  estimate both the  theoretical and experimental values of  ratio $r$ between the diamagnetic  peaks  at the main and  satellite Dirac points. According to \cite{Vallejo2021, McClure1956} the orbital magnetic susceptibility of graphene at the  main Dirac point, $\chi_M$, depends on the square of the Fermi velocity $v$ and the  disorder  standard deviation $\sigma_0$ according to:
\begin{equation}
\chi_{M}=-\frac{\sqrt{2}v^2e^2}{3\sigma_0\pi^{3/2}}
\label {chiMcClure}
\end{equation}
%Experimental and theoretical values agree within a factor 2.
 It was pointed out in \cite{Yankowitz} that the satellite Dirac cones  (at the  $m_S$ points of  the moiré hexagonal Brillouin zone) are anisotropic,  leading to an effective Fermi velocity: $\sqrt{v_{sx}\times  v_{sy}}$. 
The anisotropic components of the Fermi velocity along the principal axis of the elliptical  energy cross sections in the vicinity of   sDP are $v_{sx}= v $, the  original Fermi velocity of graphene  and $v_{sy}   =\frac{\sqrt{3}a_M t_M}{2\pi\hbar}$, the reduced transverse velocity originating from the moiré potential.  %The moiré lattice size  $a_M$ is $12.5nm$ (respectively $9.5nm$) for samples  $M_B$ (respectively $M_A$) and  the amplitude of the moiré potential
 We then  assume  that   the  susceptibility, $\chi_S$, at  secondary diamagnetic peaks,  is similar to  the McClure susceptibility of graphene in Eq.\ref{chiMcClure} replacing   $v$ by $\sqrt{v_{sx}\times  v_{sy}}$ and the amplitude of  disorder $\sigma_0$ by $\sigma_{s}$, the width of the sDP (of the order of  4.5 $\pm$  0.5 meV). This leads to the  susceptibility ratio: $ r= 3\chi_S/\chi_M= 3\frac{v_{sy}\sigma_{0}}{v\sigma_{s}}$, considering the  3 fold degeneracy of $m_s$ points. We obtain  %are $r_A=0.28$  and$ $r_B=0.57, for sample $M_A$ and %
 $r=0.55 \pm 0.1$  for sample $M_B$ to be compared to   the experimental one:  $0.33 \pm 0.1$. For these estimations we have not taken into account 
%the  contribution of Dirac points at the $\kappa_S$ points as well as 
the proximity in energy between the the sDPs  and the saddle points of bands $E_1$ and $H_1$ whose magnetic contributions of opposite sign tend to cancel each another (see SM for more details).
 %Within these approximations experimental we find that theoretical values agree within a factor 2 
%which explains the splitting experimentally observed as shown in Fig.4 on the electron side for bands E1 and E2. Larger splittings  observed on the hole side would  rather correspond to the crossings between  H2 and H3. 

Turning to the  paramagnetic susceptibility peaks, we show in SM how one can  also estimate their amplitude from the  miniband spectrum determined in our simple model  by the parameters $t_M$ and $a_M$.
%(We first note that for the parameters $t_M$ relevant for our experimental data one has only to consider the $A_1$ saddle points. )  
The DOS and the susceptibility depend on  %the  product $\alpha_^2= \alpha_x\alpha_y$  
 the curvatures $m_x^{-1}=$ and $m_y^{-1}$ of the energy bands at the saddle points. They are given in the SM for both samples and $A_1$ saddle points for the bands $E_1$, $E_2$, $H_1$, $H_2$.  These saddle points are strongly anisotropic  with   much higher curvatures along the y compared to the x  axis (along $\kappa_S, m_S$).  The ratio $ m_x/m_y$ depends strongly  on the considered bands. %as well as the moiré potential. 
As an example,  we find that it is about 20 times larger for the $E_1$, $H_1$  compared to the $E_2$, $H_2$ bands. On the other hand the geometric mean of the curvatures $\alpha = (m_x m_y)^{-1/2}$ is of the order of  $ 10 t_M a_M^2/\hbar^2 = 100 m_e^{-1}$ for all investigated  bands %E1, E1,$H_1$ and H2
%As an example,  we find that the ratio  for the E2 band which has the highest curvatures at the $A_1$ saddle points, we find $\alpha_x$ = -3.6 $\alpha_y$  = 100 $m_e^{-1}$ and   $\alpha_x$ =- 3.6 $\alpha_y$ =48  $m_e^{-1}$ for respectively sample $M_A$ and $M_B$, 
where  $m_e$ is the free electron mass. %$\alpha_x$ = 4.7 $t_M a_M^2$  and $\alpha_y$  -1.5 $t_M a_M^2$, 

In order to go further and  compare these findings to the experiments, one also needs  to take into account  the effect of  disorder. Experiments show that the paramagnetic peaks above the sDP  in the electron doped region are of the same order of magnitude as the diamagnetic ones at the sDP. We  were expecting instead smaller amplitudes due to the weaker logarithmic divergences   at the vH singularities compared to the  delta peak anomalies at sDPs. However  we show in SM that these results can be understood  when considering  that the   disorder induced rounding   of the logarithmic paramagnetic magnetisation singularities   at   saddle points  is smaller  than for the  diamagnetic ones  at the sDPs.  From the experimental values of the ratio  $ \sigma_{s}/t_M$  %estimated experimentally of the order of 0.2  and 0.33 for $M_A$ and $M_B$ 
and  the estimated  band curvatures, one finds that the maxima  of the  paramagnetic  magnetisation  at $A_1$ saddle points  for the $H_2$ and the $E_2$ bands  are   of the same order of magnitude as the diamagnetic  peaks at the sDPs,   and  larger than the  saddle point paramagnetism  on  $H_1$ and  $E_1$ bands  ( reduced by the vicinity of the SDP and barely seen experimentally for sample $M_B$). We show in SM that it is possible to  reproduce at least qualitatively, the energy positions and relative amplitudes of the two diamagnetic and paramagnetic peaks in the vicinity of the the sDPs in the electron doping range.  %Our experimental data are unfortunately not accurate enough to distinguidh between the 1/3  power law and logarithmic sigularities expected theoretically.

Finally we note that the large miniband curvatures,  due to the large period  of the  moiré potential,  contrast with the curvatures  at the saddle points of the original   2D  atomic lattice  \cite{Vignale91}  and exclude any sizable contribution of Pauli magnetism at  vH singularities. The Pauli susceptibility $\chi_P$ is indeed expected to vary as $\chi_P = \mu_B^2 \rho(\epsilon)$ where $\mu_B = e \hbar/2 m_e $ is the Bohr magneton, yielding a ratio between the    Vignale orbital and  Pauli spin  susceptibility  $\chi_{V}/\chi_P =   m_e^2/m_x m_y $   of the order of $10^{4}$, whereas this ratio is  of the order of 1 for the  2D square lattice.

 In conclusion, our measurements of the orbital  magnetisation of graphene with a  moiré potential show a rich set of  singularities of the orbital magnetisation in the vicinity of sDPs. These   consist of diamagnetic peaks at the satellite Dirac points surrounded by paramagnetic peaks which can be associated to the van Hove singularities of the DOS  at the saddle points of the  mini-band structure   induced by the moiré potential.   These experiments  therefore confirm  the long standing theoretical predictions of the existence of    paramagnetic   orbital magnetism in 2D materials at van Hove singularities which, in the case of  the graphene/hBN  moiré investigated here, exceeds by far the Pauli susceptibility. A natural prolongation of this work would be to measure the orbital magnetisation of  bilayer graphene moiré structures  also extensively investigated \cite{Andrei2010, Andreireview} with the possibility to obtain ferromagnetic orbital phases,\cite{Zeldov2022}. It is also interesting that the  typical amplitude of the paramagnetic susceptibility peaks we measure is of the same order 
of magnitude than the values predicted for graphene bilayer moirés close to the magic angle \cite{Guerci2021}. This  singular paramagnetic orbital magnetism is  shown to possibly lead to the emergence of new kinds of correlated phases  when the sample is embedded in a quantum  electromagnetic cavity. Our  results also motivate the  extension of this work to graphene twisted bilayers with larger moiré periods in which field periodic orbital currents  are expected \cite{Geim2019}.  

Aknowledgements: The authors thank E. Paul of SPEC-CEA for the GMR sensors patterning R.Weil  and S. Autier-Laurent   of LPS for Nanofabrication and Cryogenic support.
They also aknowledge fruitfull discussions with  A.Chepelianskii,  F. Parmentier, R.Delagrange, D. Mailly, C. Mora, P.Simon and financial support from the   BALLISTOP ERC 66566 advanced grant.
This work benefited  from  the C2N micro nanotechnologies platforms and partly supported by the RENATECH network, the General Council of Essonne and the DIM-SIRTEC. R.R.-P. acknowledge the ERC starting grant TWISTRONICS. K.W. and T.T. acknowledge support from JSPS KAKENHI (Grant Numbers 19H05790, 20H00354 and 21H05233).


\begin{thebibliography}{99}
\bibitem{notedetaVg} For samples $M_A$ the position in  gate voltage of the diamagnetic peaks differ by  2 volts  compared to the resistance peaks. This offset which corresponds to a $5\%$  shift in chemical potential can be  attributed to the small size of the gate compared to the size of the sample measured in transport.
\end{thebibliography}
\end{document}

\today

\title{Supplementary material for: Paramagnetic  singularities of the orbital magnetism in graphene with a moir\'e potential}
\author{J. Vallejo Bustamante}
\affiliation{Universit\'e Paris-Saclay, CNRS, Laboratoire de Physique des Solides, 91405  Orsay, France.}
\author{R. Ribeiro-Palau}
\affiliation{Université Paris-Saclay, CNRS, C2N, 91120 Palaiseau, France.}
\author{C. Fermon}
\author{M. Pannetier-Lecoeur}
\affiliation{SPEC, CEA, CNRS, Universit\'e Paris-Saclay, 91191 Gif-sur-Yvette, France.}
\author{ K. Watanabe}
\affiliation{Research Center for Functional Materials, National Institute for Materials Science, 1-1 Namiki, Tsukuba 305-0044, Japan}
\author{T. Tanigushi}
\affiliation{International Center for Materials Nanoarchitectonics, National Institute for Materials Science,  1-1 Namiki, Tsukuba 305-0044, Japan}
\author{R. Deblock}
\author{S. Gu\'eron}
\author{M. Ferrier}
\affiliation{Universit\'e Paris-Saclay, CNRS, Laboratoire de Physique des Solides, 91405  Orsay, France.}
\author{J.N. Fuchs}
\affiliation{Sorbonne Universit\'e, CNRS, Laboratoire de Physique Théorique de la Mati\`ere Condens\'ee, LPTMC, 75005 Paris, France}
\author{G. Montambaux}
\author{F. Pi\'echon}
\author{H. Bouchiat}
\affiliation{Universit\'e Paris-Saclay, CNRS, Laboratoire de Physique des Solides, 91405  Orsay, France.}

\maketitle

\section{Mini-band structure of graphene with a moir\'e potential}
\subsection{Band structure}	
\begin{figure}[h!]
\includegraphics[scale=0.5]{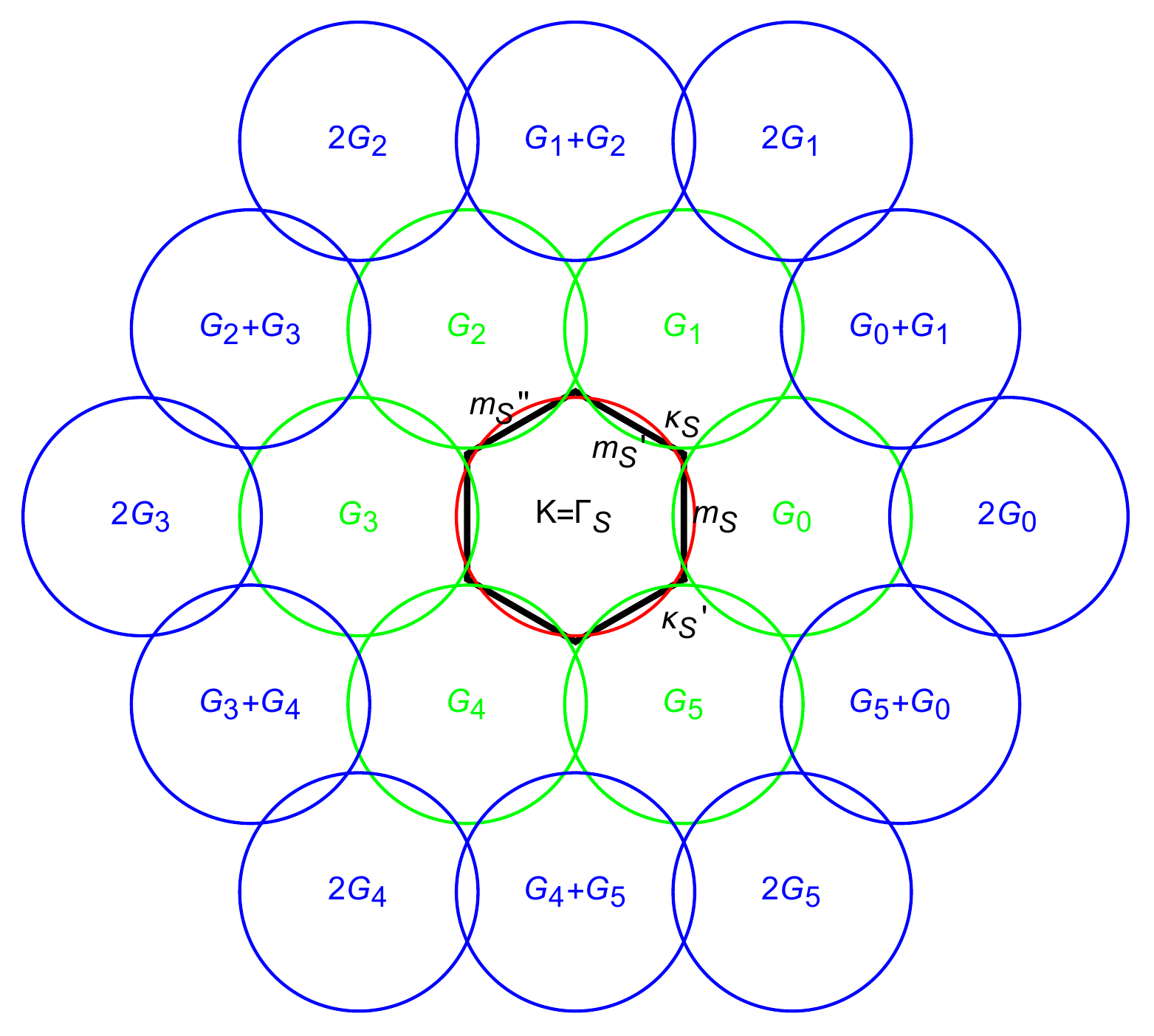}
\caption{Truncation in reciprocal space showing a central Dirac cone section (red circle of radius $0.55G$), surrounded by a first shell of 6 green circles and a second shell of 12 blue circles. The mBZ is shown as a black hexagon and high symmetry points are indicated ($\Gamma_S$, $\kappa_S$ with 3-fold degeneracy (C3 symmetry) and $m_S$ with 2-fold degeneracy (C2 degeneracy)). In the limit of vanishing $t_M$, there is a Dirac point at $m_S$ between the two lowest electron bands and between the two lowest hole bands. Correspondingly at $\kappa_S$, there is a 3-fold degeneracy between the three lowest electron bands and between the three lowest hole bands. Dirac points emerge at the three $m_S$ points and unusual $C_3$ saddle points appear at the two $\kappa_S$ points.}
\label{19cones}
\end{figure}

In order to investigate the orbital susceptibility of  the graphene layer  with a moir\'e potential on the graphene , we develop the approach of Yankowitz et al~\cite{Yankowitz} (see also Wallbank et al~\cite{Wallbank})  to determine the mini-band structure. We start from a 2D massless Dirac Hamiltonian $H_0$ representing the unperturbed graphene layer in a single valley (say the $K$ valley) and add a scalar periodic potential $V$ with 6-fold symmetry to represent the effect of the moir\'e. The total Hamiltonian reads
\beq
H = H_0+V = v \boldsymbol{p}\cdot \boldsymbol{\sigma} + \sigma_0 t_M \sum_{m=0}^5 e^{i \boldsymbol{G}_m \cdot \boldsymbol{r}},
%= v \boldsymbol{p}\cdot \boldsymbol{\sigma} + t_M \sum_{m=0}^5 \cos (\boldsymbol{G}_m \cdot \boldsymbol{r}), 
\eeq	
where $\boldsymbol{\sigma} $ are the Pauli matrices describing the sublattice pseudospin of the honeycomb lattice, $\sigma_0$ is the $2\times 2$ unit matrix and $v$ is the Fermi velocity of graphene ($v\sim 10^6$~m/s). The moir\'e  potential has an amplitude $t_M$ and depends on reciprocal lattice vectors $\boldsymbol{G}_m=G\{\cos(m \pi/3), \sin(m \pi/3)\}$ with norm $G=\frac{4\pi}{\sqrt{3}a_M}$, where $a_M$ is the  moiré-lattice period. The two vectors $\boldsymbol{G}_1$ and $\boldsymbol{G}_2$ form a basis for the reciprocal lattice (see Fig.~\ref{19cones}). A negative (resp. positive) $t_M$ means that the potential minima form a triangular (resp. honeycomb) lattice. In the present experiment, $a_M$ is of the order of 10 nm and $t_M$ is estimated to be $\sim -25$~meV. %Yankowitz et al. theoretically estimate $t_M=30$~meV (they give $V=2t_M=60$~meV)~\cite{Yankowitz,Wallbank}. 
The momentum operator is the usual $\boldsymbol{p}\to -i \hbar \boldsymbol{\nabla}$ in 2D (but it is shifted such that $K$ now plays the role of the reciprocal space origin $\Gamma_S$). We set $\hbar\equiv 1$, $v\equiv 1$ and take $G^{-1}=\frac{\sqrt{3}}{4\pi}a_M\sim 1$~nm as unit length. Therefore, $\hbar v G\simeq 0.4$~eV is the unit energy and the only remaining dimensionless parameter is $t_M\sim -0.06$.

Eigenvectors of $H_0$ are plane wave spinors $|\boldsymbol{k},s\rangle$ where $s=\pm$ is the sign of the energy $\varepsilon_{\boldsymbol{k},s}^{(0)}=s k$ and $\boldsymbol{k}$ is a wavector.  We write $H$ in the eigenbasis of $H_0$. The potential has matrix elements:
\beq
\langle \boldsymbol{k}',s'|V|\boldsymbol{k},s\rangle = t_M \delta_{s,s'} \sum_m \delta_{\boldsymbol{k}',\boldsymbol{k}+\boldsymbol{G}_m} .
\eeq
As we are only interested in the first few bands (typically the 6 bands closest to zero energy), we truncate the Hilbert space at low energy. For each $\boldsymbol{k}$ in the first Brillouin zone (called moir\'e or mini Brillouin zone, mBZ, see Fig.~\ref{19cones}), we only keep states $|\boldsymbol{k}',s\rangle$ such that $\boldsymbol{k}'-\boldsymbol{k}$ is a reciprocal lattice vector (i.e. $\boldsymbol{k}'-\boldsymbol{k}=c_1 \boldsymbol{G}_1 +  c_2 \boldsymbol{G}_2$ with $c_1$ and $c_2$ integers) such that $|\boldsymbol{k}'-\boldsymbol{k}|\leq 2$. This means that the energy $|\varepsilon|$ should be smaller than the cutoff $\varepsilon_c=2$. At positive energy, there are 19  such states, see Fig.~\ref{19cones} showing one central Dirac cone circle in red (of radius 0.55 chosen close to the satellite Dirac points and such as to show intersections between the circles), 6 nearest neighbors in green (first shell) and 12 next-nearest neighbors in blue (second shell). 

\begin{figure}[h!]
\includegraphics[scale=0.35]{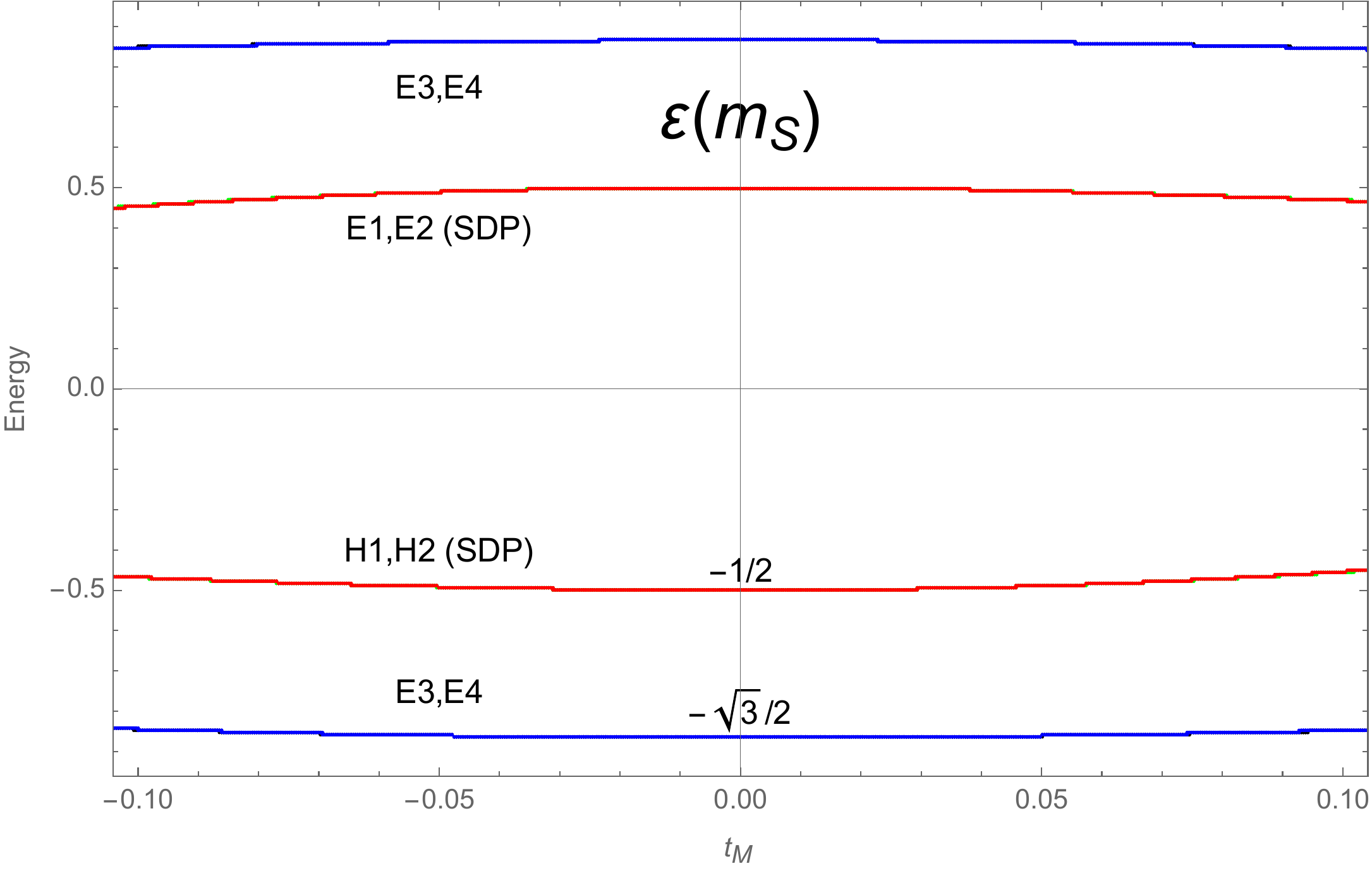}
\includegraphics[scale=0.35]{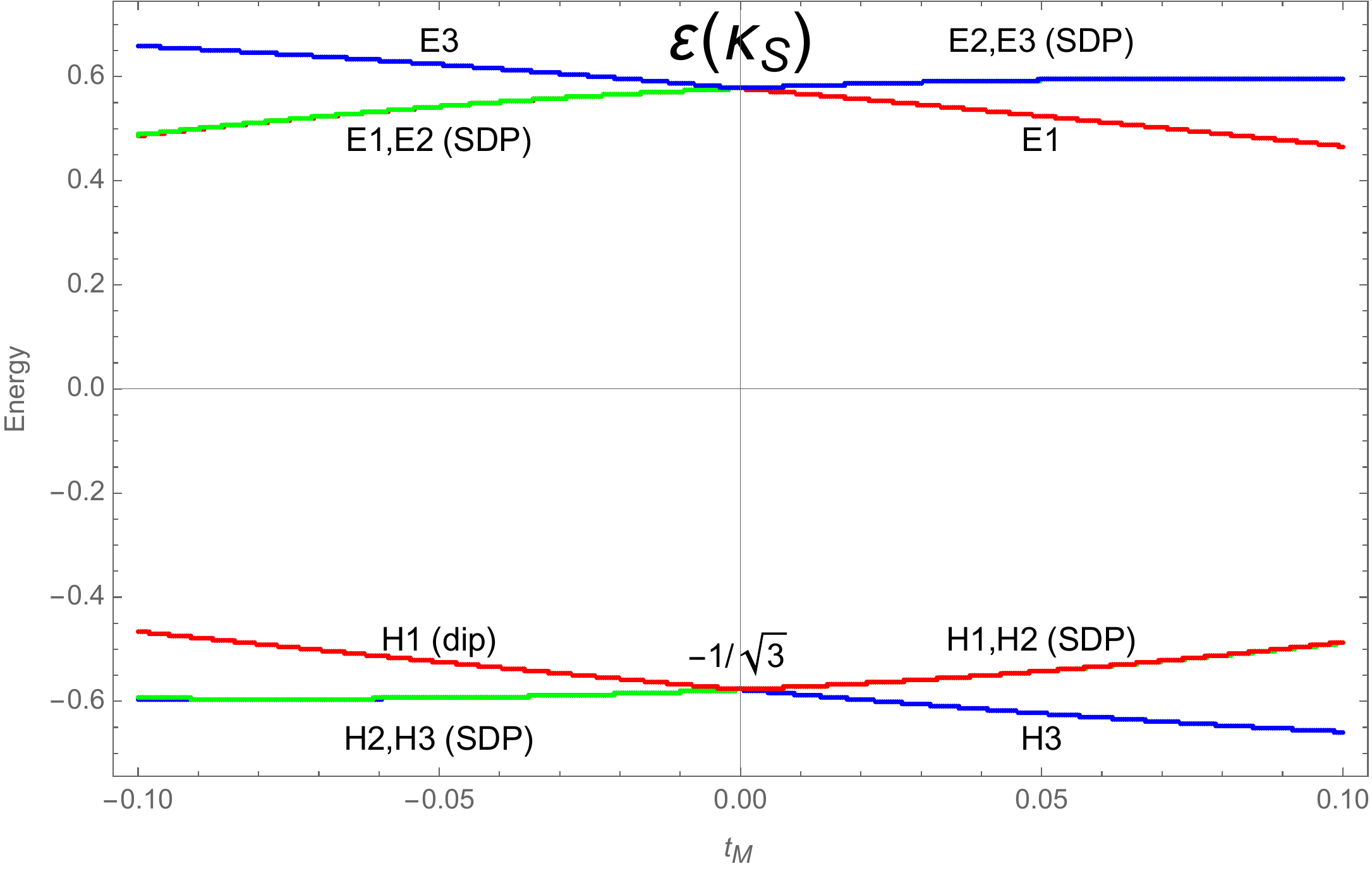}
\includegraphics[scale=0.4]{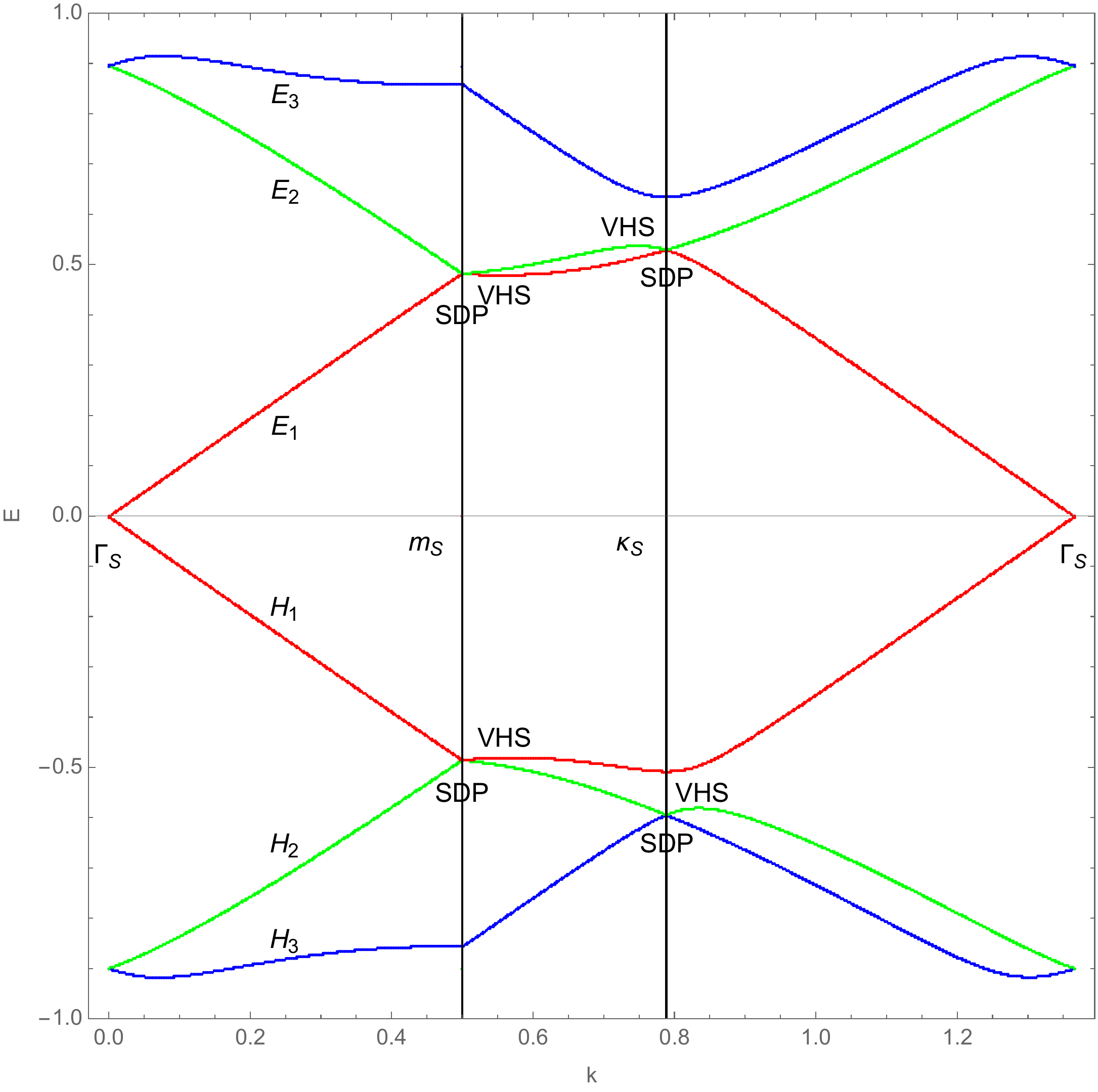}
\caption{ Upper curves: evolution of the low-energy energy bands at  $m_S$, left  and  $\kappa_S$  right as a function of the strength of the moir\'e potential $t_M$ (in units of $\hbar v G)$. Lower curves:band structure along the closed path $\Gamma_S \to m_S \to \kappa_S \to \Gamma_S$ in reciprocal space ($k$ is a wavevector in units of $G$) showing the 3 upper hole bands ($H_1$, $H_2$, and $H_3$) and 3 lower electron bands ($E_1$, $E_2$, and $E_3$) for graphene in the presence of a $C_6$ moir\'e potential. The moir\'e potential has a strength $t_M=-0.064$.}
\label{figs1}
\end{figure}
For each wavevector in the mBZ, we numerically diagonalize the Hamiltonian matrix to obtain 38 bands, of which we only keep the 6 closest to zero energy (3 at positive and 3 at negative energy). As $|t_M| \ll 1$, we are in a regime of nearly-free Dirac electron and the free band structure $\varepsilon_{\boldsymbol{k},s}^{(0)}=s k$ should mostly be affected near mBZ edges. Because of truncation in reciprocal space, our calculation is only valid for $|t_M| \ll  1$, which means that we will restrict to $|t_M|\leq 0.1$. 

\begin{figure}[h!]
\includegraphics[width=0.6\linewidth]{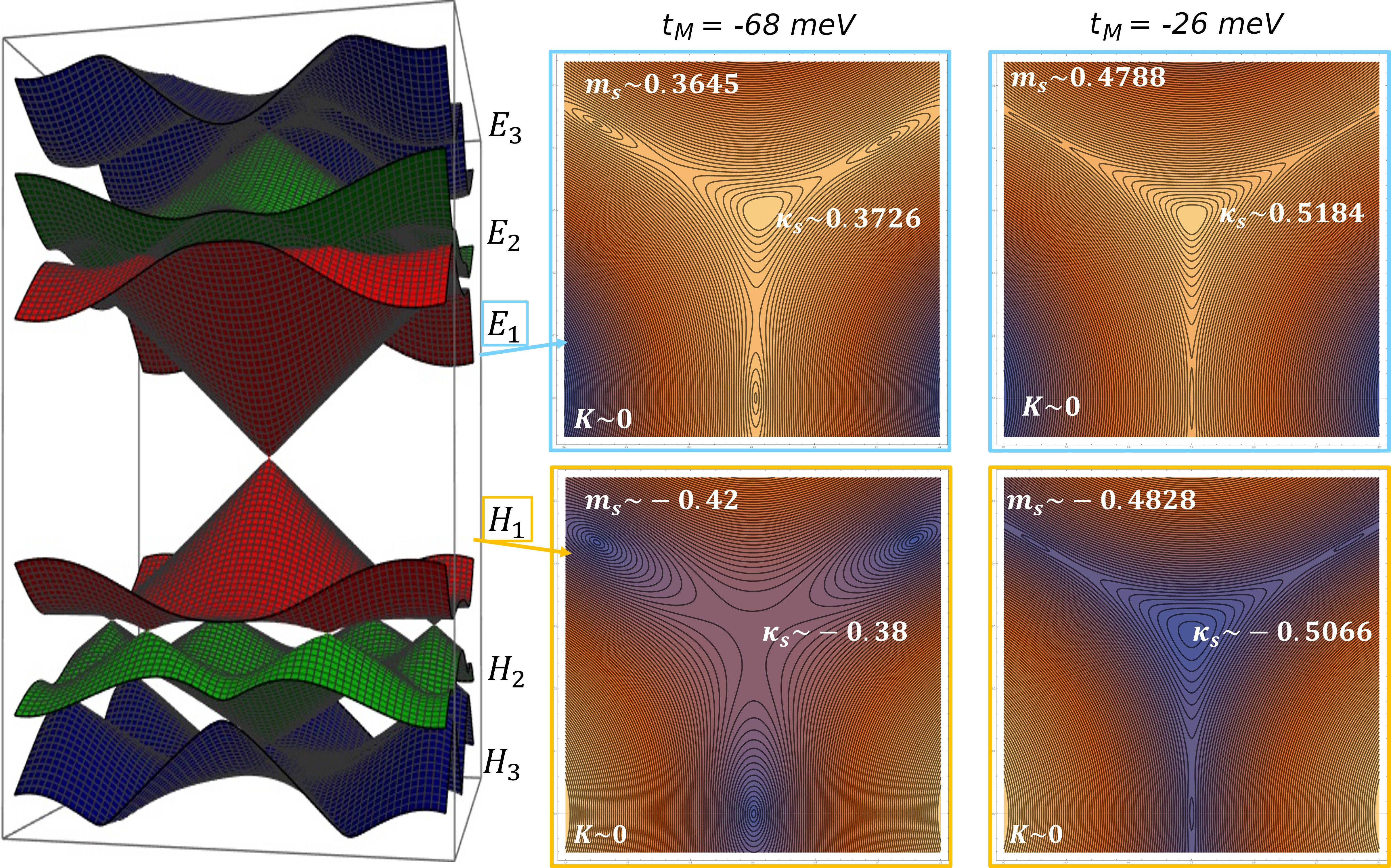}
\caption{(Left) 3D plot showing the 3 upper hole bands ($H_1$, $H_2$, and $H_3$) and 3 lower electron bands ($E_1$, $E_2$, and $E_3$) for graphene in the presence of a $C_6$ moir\'e potential. (Right) Contour plots of the highest hole band $H_1$ and lowest electron band $E_1$ for a moir\'e potential $t_M=-0.17\simeq -68$ meV (in units of $\hbar v G\sim 0.4$~eV). For $H_1$, one can notice sDPs at the 3 $m_S$ points and an unusual $C_3$ saddle point at each $\kappa_S$ point.  At a lower value $t_M = -0.064\simeq -26$ meV, a dip appears at $\kappa_S$ in $H_1$ resulting in the formation of 6 usual $A_1$ saddle points in between $\kappa_S$ and $m_S$. For $E_1$ and for both values of $t_M$, there are 3 sDPs at $m_S$, 2 sDPs at $\kappa_S$ and 6 $A_1$ saddle points.}\label{figs3}
\end{figure}

As there is a symmetry between positive and negative energy upon changing the sign of $t_M$ (see Fig.~\ref{figs1}, we only discuss $t_M\leq 0$ (typically $t_M\sim -0.05$). 
We find that, on top of the main Dirac point (crossing of $E_1$ and $H_1$, at zero energy and $K=\Gamma_S$), the mini band structure consists of 3 satellite Dirac points (sDP) at $m_S$ between the two lowest bands $E_1$ and $E_2$ (at energy $\varepsilon\approx 0.5$) and 2 extra ones at $\kappa_S$ between $E_1$ and $E_2$ (at energy $\varepsilon\lesssim 1/\sqrt{3}\approx 0.57$). The two sets of sDP are connected by 6 (in each band) usual $A_1$ saddle points between $m_S$ and $\kappa_S$, see Figs.~\ref{figs1} and \ref{figs3}. 

The situation is more subtle on the hole side: there are 3 sDPs at $m_S$ between the two lowest bands $H_1$ and $H_2$ (at energy $\varepsilon\approx -0.5$) and 2 extra ones at $\kappa_S$ between $H_2$ and $H_3$ (at energy $\varepsilon\lesssim -0.57$). The band $H_1$ does not touch $H_2$, but it features a dip (with $C_3$ symmetry) at $\kappa_S$. There are also 6 $A_1$ saddle points between $m_S$ and $\kappa_S$, see Figs.~\ref{figs1} and \ref{figs3}(A,C).
The existence of two sets of sDP at nearby energy ($\sim 0.5$ and $\sim 0.57$) may explain the observed splitting of the McClure diamagnetic peak. This splitting appears larger on the hole side than on the electron side and may be explained by the fact that the two sets of sDP are between $H_1$ and $H_2$ at $m_S$ but between $H_2$ and $H_3$ at $\kappa_S$.

If we increase $t_M$ up to $-0.17$ (a value at which our calculations start to suffer from the truncation), we find that three $A_1$ saddle points merge into an unusual $C_3$ saddle point at $\kappa_S$ in the $H_1$ band, see Fig~\ref{figs3}(B). This $C_3$ saddle point seems to exist for $t_M$ in between $-0.14$ and $-0.25$. There are two-fold degeneracies at $m_S$ and $\kappa_S$ that are never split by $t_M$ and are responsible for the existence of Dirac points.
\begin{figure}
\centering
    \includegraphics[width=0.5\linewidth]{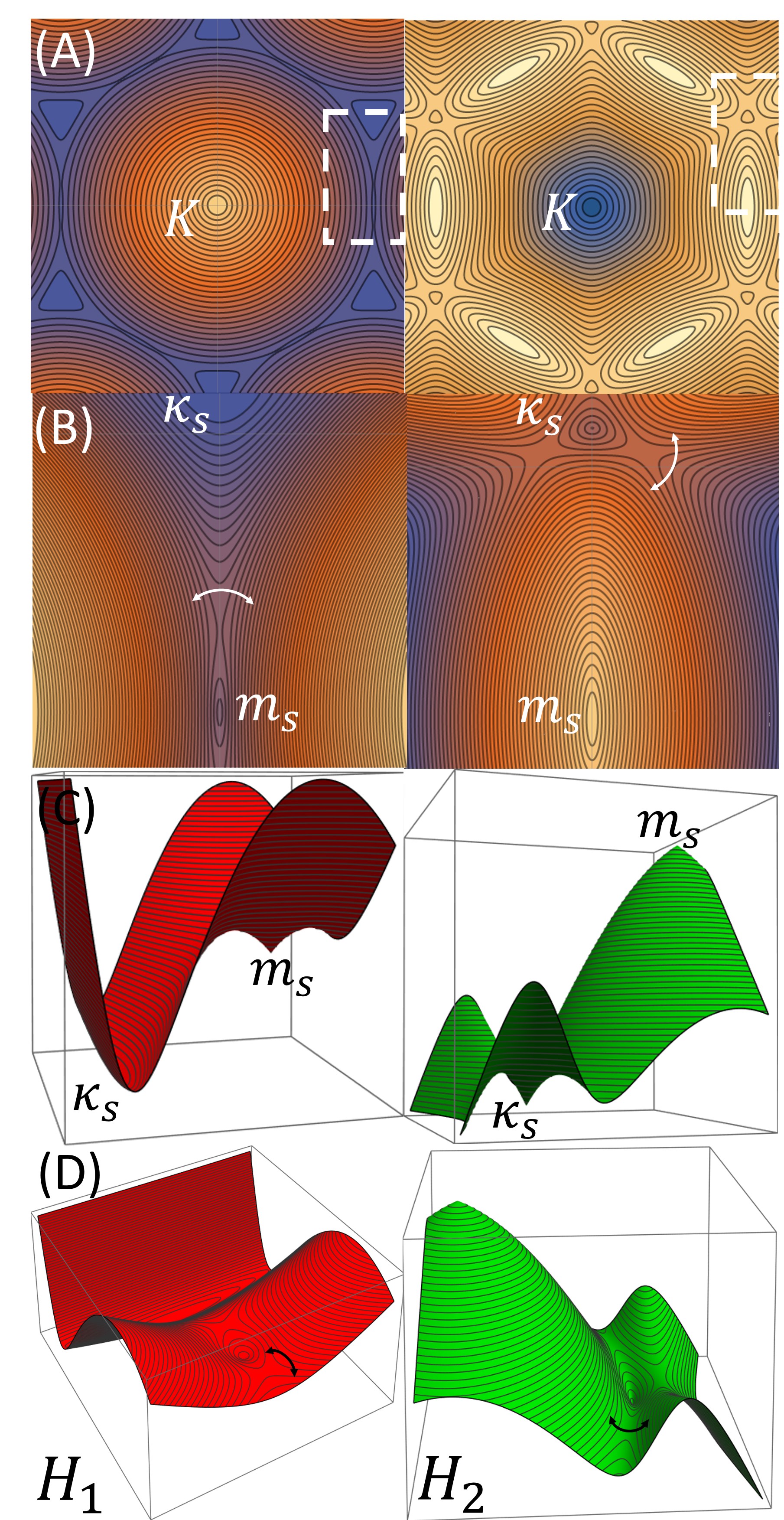}
    \caption{(A) Contour plot of bands $H_2$ (left) and $H_1$ (right) in the mini Brillouin zone. (B) Zoom around $\kappa_s$ and $m_s$ points. (C) and (D) 3D representation of the bands close to $\kappa_s$ and $m_s$. The curved double arrow indicates the saddle point.}
    \label{fig:bandesHoles}
\end{figure}

\begin{figure}
\centering
    \includegraphics[width=0.5\linewidth]{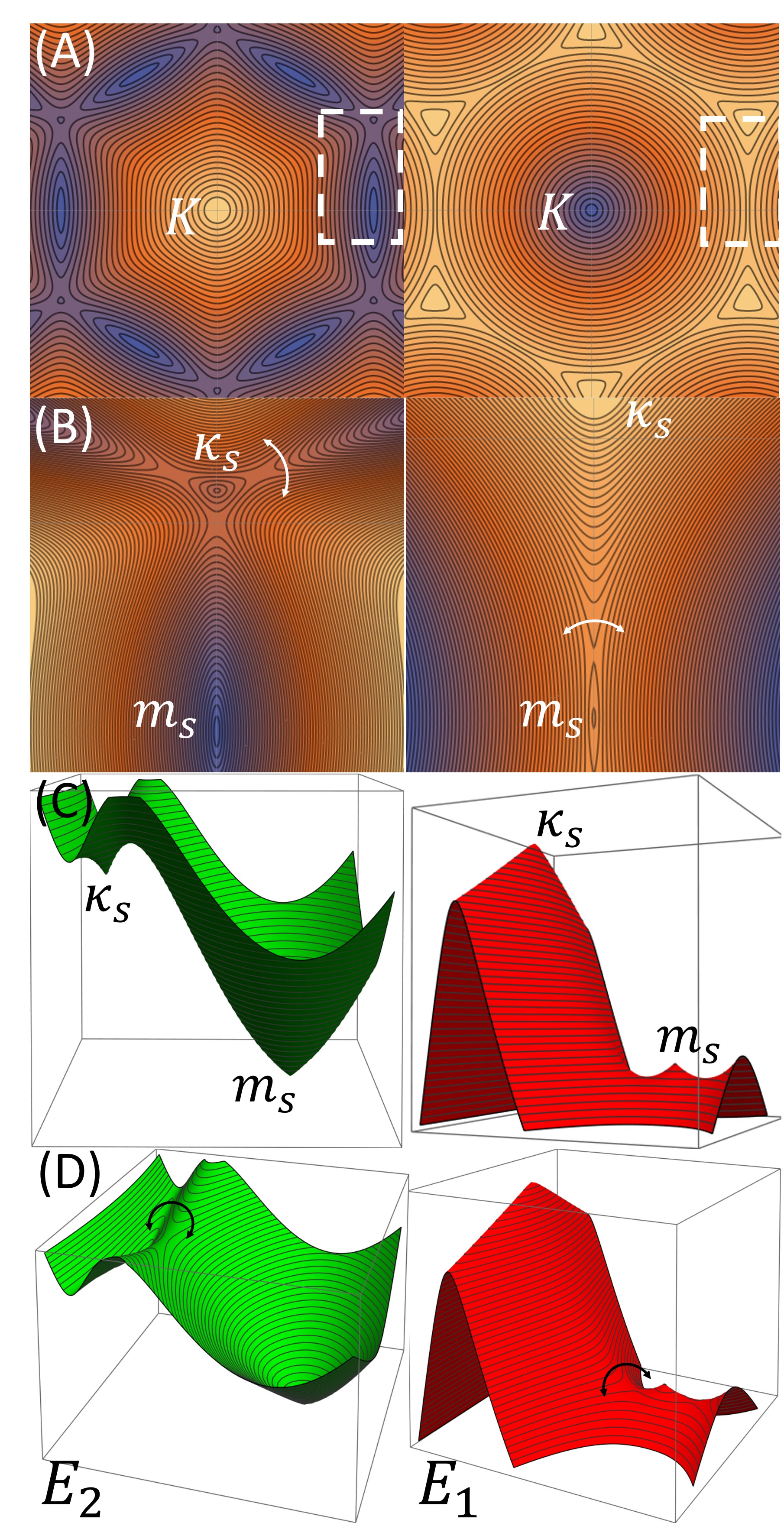}
    \caption{(A) Contour plot of bands $E_1$ (left) and $E_2$ (right) in the mini Brillouin zone. (B) Zoom around $\kappa_s$ and $m_s$ points. (C) and (D) 3D representation of the bands close to $\kappa_s$ and $m_s$. The curved double arrow indicates the saddle point.}
    \label{fig:bandesElec}
\end{figure}

We show in Fig.\ref{fig:bandesHoles}  and Fig.\ref{fig:bandesElec} contour plots for the different bands  $E_1,$ $E_2$ and  $H_1,$ $H_2$ calculated for $t_M$ =-23 meV,  zooming around $\kappa_S$ and $m_S$. This data illustrates the strong anisotropy of the saddle points on the $H_1$ and $E_1$ bands compared to the $H_2$ and $E_2$ bands. We note that the very high estimated values  of the  curvatures $\alpha_x= 1/m_x$ and $\alpha_y=1/m_y$  for the $E_1$ and $H_1$ bands are only meaningfull in a very small energy range beyond which the bands vary linearly with k.  These band structures are expected to be  rounded  by disorder, leading to smaller effective values of $\alpha_y$  and therefore $\alpha$.

\subsection{Analytical approximation}
In this section, we construct effective minimal Hamiltonian models near the $m_S$ and $\kappa_S$ points in the presence of the moir\'e coupling potential and explore the nature of the saddle points of the first two electron bands $E_{1,2}$ (resp. hole bands $H_{1,2}$)  along the line joining $m_S$ to $\kappa_S$.

Considering the low energy Dirac Hamiltonian $H({\bm k})=k_x \sigma_x +k_y \sigma_y$, the corresponding pseudospin eigenstates 
$|s,{\bm k}\rangle$ of energy $\varepsilon_{{\bm k},s}^{(0)}=s |{\bm k}|$
may be written as
\begin{equation}
|s,{\bm k}\rangle=\frac{1}{\sqrt{2}}\left(
 \begin{array}{c}
  s\\
  \frac{k_x+i k_y}{|{\bm k}|}
 \end{array}
\right).
\end{equation}
We can then obtain the overlap matrix between the two eigenstates $|s,{\bm k}\rangle$ and $|s,{\bm k'}\rangle$
\begin{equation}
\langle s,{\bm k}|s,{\bm k'}\rangle=\frac{1}{2}(1+\frac{{\bm k}\cdot{\bm k'}}{|{\bm k}||{\bm k'}|}+i\frac{{\bm k}\times{\bm k'}}{|{\bm k}||{\bm k'}|}).
\end{equation}
Hereafter, since the overlap does not depend on the band index we simply denote $\langle s,{\bm k}|s,{\bm k}'\rangle\equiv \langle {\bm k}|{\bm k}'\rangle$.
For what follows what is needed is the overlap matrix between
 ${\bm k}_1=k {\bm e}_{1}+{\bm q}$ and ${\bm k}_2=k{\bm e}_2+{\bm q}$ with $ {\bm e}_{1,2}$ two distinct unit vectors. To order ${\bm q}$ one finds,
\begin{equation}
\langle k{\bm e}_1+{\bm q}|k{\bm e}_2+{\bm q}\rangle=
\frac{1}{2}(1+({\bm e}_1\cdot{\bm e}_2+i {\bm e}_1\times{\bm e}_2)(1-\frac{ {\bm q}\cdot ({\bm e}_1+{\bm e}_2)}{k})+ \frac{ {\bm q}\cdot ({\bm e}_1+{\bm e}_2)}{k}
+i\frac{ {\bm q}\times ({\bm e}_1-{\bm e}_2)}{k});
\label{eq:overlap}
\end{equation}
where to order ${\bm q}^2$ one has
\begin{equation}
 |k {\bm e}_{1}+{\bm q}|=\sqrt{(k^2+{\bm q}^2+2k{\bm q}\cdot{\bm e}_1)}=k +{\bm q}\cdot{\bm e}_1+\frac{ ({\bm q}\times{\bm e}_1)^2}{2k}.
 \label{eq:norm}
\end{equation}

%As in JN notes, the reciprocal lattice vectors of the moir\'e lattice are ${\bf G}_m=({\cos(m\frac{\pi}{3}), \sin(m\frac{\pi}{3})})$ in unit of $G = \frac{4\pi}{\sqrt{3} a_m}$. Hereafter all wavevectors variables are dimensionless quantities written in unit of $G$. The moir\'e potential strength  is noted $t_M \ll 1$ in unit of $E_G=\hbar v_F G$.\\

\subsubsection{Effective Hamiltonian around $m_S$}
According to Fig.~\ref{19cones}, the two bare/primary Dirac points closest to the point $m_S$ are situated at $\Gamma_S$ and ${\bf G}_0$ and verify ${\bm k}_{m_S}=\Gamma_S+{\bm k}_+={\bf G}_0+{\bm k}_-$ 
with ${\bm k}_{\pm}=\pm k{\bm e}_x$ where $k=\frac{1}{2}$ in units of $G$.
Before turning on the moir\'e potential, around $m_S$, the two degenerate electron bands $E_{1,2}=k=\frac{1}{2}$ (resp. hole bands $H_{1,2}=-k$) 
correspond to the states $|s,{\bm k}_{\pm}\rangle$ with $s=+$ for electron and $s=-$ for hole.

Turning on the moir\'e potential, the effective Hamiltonian around  $m_S$ that only considers the coupling between the degenerate bands reads
(with $|{\bm q}| \ll 1$)
\begin{equation}
{\cal H}_{m_S}=\left( 
\begin{array}{cc}
s |{\bm k}_{+}+{\bm q}|& t_M \langle {\bm k}_++{\bm q}|{\bm k}_{-}+{\bm q}\rangle\\
t_M\langle {\bm k}_{-}+{\bm q}|{\bm k}_{+}+{\bm q}\rangle& s |{\bm k}_{-}+{\bm q}|\\
 \end{array}\right),
\end{equation}
To correctly describe the energy spectrum at order ${\bm q}^2$ it is sufficient to take the approximate expressions Eqs.~(\ref{eq:overlap}, \ref{eq:norm}), we obtain:
\begin{equation}
{\cal H}_{m_S}=s(k+\frac{q_y^2}{2k}) \sigma_0 + s q_x \sigma_z -\frac{t_M}{k} q_y  \sigma_y .
\end{equation}
The last two terms of the above Hamiltonian correspond to an anisotropic Dirac cone at $m_S$ with effective velocities $c_x=|s|=1$ and $c_y=|t_M|/k=2|t_M|$.

For $s>0$ (electron side), one obtains the band dispersions
$E_1({\bm q})=k +\frac{q_y^2}{2k} -\sqrt{q_x^2+\frac{t_M^2}{k^2} q_y^2 }$
and $E_2({\bm q})=k +\frac{q_y^2}{2k} +\sqrt{q_x^2+\frac{t_M^2}{k^2} q_y^2 }$. 
The corresponding energy spectrum resembles that of a Rashba Hamiltonian however with an effective strong anisotropy in the Rashba-like coupling. 
This anisotropy in the Rashba-like coupling appears  to be at the origin of the presence of two saddle points in the band  $E_1({\bm q})$ whereas it is obvious that the band $E_2({\bm q})$ has no saddle point.
More quantitatively the band $E_1({\bm q})$ exhibits a pair of saddle points at postions ${\bm q}^*=(0,\pm t_M)$ which, as expected, are along the line joining $m_S$ to $\kappa_S$.
Expanding $E_1({\bm q})$ to quadratic order around the saddle points one finds
\beq
E_1({\bm q}^*+{\bm q})=k+\frac{(t_M+q_y)^2}{2k}-\sqrt{q_x^2+\frac{t_M^2}{k^2} (t_M+q_y)^2}\simeq E_1^*+\frac{q_y^2 }{2m_{yy}}+\frac{q_x^2 }{2m_{xx}}
\eeq
with $E_1^*=k-\frac{t_M^2}{2k}$, $m_{yy}=k$ and $m_{xx}=-\frac{t_M^2}{k}$. The corresponding curvatures are $\alpha_x=1/(2 m_{xx})=-1/(2 t_M)^2$ and $\alpha_y=1/(2 m_{yy})=1$. 
From these expressions, one deduces that the saddle points energy is shifted by an amount $-\frac{t_M^2}{2k}$ from the secondary Dirac point at $m_S$.
This energy shift, as well as the negative effective mass $m_{xx}$, are very small since they are quadratic in the moir\'e potential strength $t_M$. This also means that the saddle point is very anisotropic ($|\alpha_x/\alpha_y|=1/(2 t_M)^2 \sim 100$).

For hole bands, since $H_{1,2}({\bm q})=-E_{1,2}({\bm q})$, one deduces that only $H_1({\bm q})$
exhibits saddle points (at positions ${\bm q}^*=(0,\pm t_M)$) with energy shift and effective masses opposite to that of $E_1({\bm q})$.

\subsubsection{Effective Hamiltonian around $\kappa_S$}
According to  Fig.~\ref{19cones}, the three bare Dirac points closest to  $\kappa_S$ are situated at positions $\Gamma_S, {\bf G}_0$ and ${\bf G}_1$ that verify
${\bm k}_{\kappa_S}=\Gamma_S +k{\bm e}_1= {\bf G}_0+k{\bm e}_2= {\bf G}_1 +k{\bm e}_3$,
with $k=\frac{1}{\sqrt{3}}$ and ${\bm e}_1=(\frac{\sqrt{3}}{2},\frac{1}{2})$ ,
${\bm e}_2=(-\frac{\sqrt{3}}{2},\frac{1}{2})$ and  ${\bm e}_3=(0,-1)$ such that $\sum_i {\bm e}_i=0$. 
Before turning on the moir\'e potential there is a three fold degeneracy in bands $E_{1,2,3}=k=\frac{1}{\sqrt{3}}$ (resp. $H_{1,2,3}$) at each $\kappa_S$ point.
These three fold degenerate bands correspond to the states $|s,k{\bm e}_{1,2,3}\rangle$ with $s=+$ for electron and $s=-$ for hole.
Turning on the moir\'e potential, the effective Hamiltonian around the point $\kappa_S$ that only considers the coupling between the degenerate bands reads
(with $|{\bm q}| \ll 1$)
\begin{equation}
{\cal H}_{\kappa_S}=\left( 
\begin{array}{ccc}
s |k{\bm e}_{1}+{\bm q}|& t_M \langle k{\bm e}_1+{\bm q}|k{\bm e}_{2}+{\bm q}\rangle&t_M \langle k{\bm e}_1+{\bm q}|k{\bm e}_{3}+{\bm q}\rangle\\
t_M\langle k{\bm e}_{2}+{\bm q}|k{\bm e}_{1}+{\bm q}\rangle& s |k{\bm e}_{2}+{\bm q}|&t_M\langle k{\bm e}_{2}+{\bm q}|k{\bm e}_{3}+{\bm q}\rangle\\
t_M\langle k{\bm e}_{3}+{\bm q}|k{\bm e}_{1}+{\bm q}\rangle&t_M\langle k{\bm e}_{3}+{\bm q}|s,k{\bm e}_{2}+{\bm q}\rangle& s |k{\bm e}_{3}+{\bm q}|
 \end{array}\right).
\end{equation}
%The above form of the Hamiltonian around $\kappa_S$ point verfies a $C_3$ symmetry. As a consequence, up to order ${\bm q^2}$ the energy spectrum
%is fully isotropic around $\kappa_S$ and therefore it cannot exhibits any saddle point to this order. The first anisotropic terms are of order $q_x^2q_y$ but appear insufficient to correctly 
%describe the energy spectrum. To correctly describe the dispersion relation of energy bands around $\kappa_S$ it is necessary to take the exact form (to all order in ${\bm q}$)
%of both the off diagonal and diagonal elements of ${\cal H}_{\kappa_S}$. Interestingly, this simple $3\times3$ Hamiltonian appears to describe correctly 
%the behavior around the three $m_S$ points nearest to $\kappa_S$; it also automatically recovers the primary Dirac points for $E_{1}$ and $H_{1}$ at $\Gamma_S,{\bm G}_{0,1}$ 
%which corresponds to ${\bm q}=-k {\bm e}_{1,2,3}$ respectively. 
The above form of the Hamiltonian around $\kappa_S$ point verifies a $C_3$ symmetry. As a consequence, up to order ${\bm q^2}$ the energy spectrum
is fully isotropic around $\kappa_S$ and therefore it cannot exhibits any saddle point to this order. 
Defining $C_1(\bm q)=\textrm{Tr}{{\cal H}_{\kappa_S}}/3$, $C_2(\bm q)=\textrm{Tr}{({\cal H}_{\kappa_S}-C_1 \textrm{Id})^2}$ and $C_3(\bm q)=\textrm{Tr}{({\cal H}_{\kappa_S}-C_1 \textrm{Id})^3}$,
the three eigenenergy bands are
\begin{equation}
\lambda_{n=1,2,3}(\bm q)=C_1+\sqrt{\frac{2C_2}{3} }\cos [\frac{1}{3}\arccos{\frac{\sqrt{6}C_3}{\sqrt{C_2^3}}}+ \frac{2n\pi}{3}].
\end{equation}
More quantitatively, to order $\bm q^3$, one finds
\begin{equation}
\begin{array}{l}
C_1(\bm q)=s(k+\frac{1}{4k}\bm q^2+\frac{1}{8k^2}q_y(3q_x^2-q_y^2)),\\
C_2(\bm q)=\frac{3 t_M^2}{2}+\frac{3(4k^2+3t_M^2)}{8k^2}\bm q^2-\frac{3(4k^2-9t_M^2)}{16k^3} q_y(3q_x^2-q_y^2), \\
C_3(\bm q)=-\frac{3 t_M^3}{4}+\frac{27t_M^2(2sk+t_M)}{16k^2}\bm q^2+\frac{3(8sk^3+27t_M^3)}{32k^3}q_y (3q_x^2-q_y^2).
 \end{array}
\end{equation}
Using these expressions, for $t_M<0$, the three electron bands $E_1<E_2<E_3$ read
\begin{equation}
\begin{array}{l}
E_1(\bm q)=X(\bm q)-\frac{1}{2}[Y_+(\bm q)+\sqrt{Z_+(\bm q)}],\\
E_2(\bm q)=X(\bm q)-\frac{1}{2}[Y_+(\bm q)-\sqrt{Z_+(\bm q)}],\\
E_3(\bm q)=X(\bm q)+Y_+(\bm q).
 \end{array}
\end{equation}
Similarily for the hole bands $H_1>H_2>H_3$ one obtains,
\begin{equation}
\begin{array}{l}
H_1(\bm q)=-X(\bm q)+Y_-(\bm q),\\
H_2(\bm q)=-X(\bm q)-\frac{1}{2}[Y_-(\bm q)+\sqrt{Z_-(\bm q)}],\\
H_3(\bm q)=-X(\bm q)-\frac{1}{2}[Y_-(\bm q)-\sqrt{Z_-(\bm q)}],
 \end{array}
\end{equation}
where
\begin{equation}
\begin{array}{l}
X(\bm q)=k+\frac{1}{4k}\bm q^2+\frac{1}{8k^2}q_y(3q_x^2-q_y^2),\\
Y_\pm(\bm q)=-t_M-\frac{(2k\mp 3t_M)}{6t_M k}\bm q^2+\frac{(3t_M\pm 2k)}{18t_M^2k} q_y(3q_x^2-q_y^2), \\
Z_\pm(\bm q)=\frac{(3t_M\pm 2k)^2}{4k^2}\bm q^2+\frac{(81t_M^3-12k^2t_M\pm16k^3)}{24k^3t_M}q_y (3q_x^2-q_y^2).
 \end{array}
\end{equation}
 
To understand the content of these expressions, it is convenient to consider the limit of very small $t_M$. 
In this limit, rewriting $\tilde{q}_{x,y}=q_{x,y}/t_M $ with $|\tilde{q}_{x,y}| \le 1$ and keeping terms at most
of linear order in $t_M$ in the energy bands we can use the simplified expressions
\begin{equation}
\begin{array}{l}
X(\tilde{\boldsymbol{q}})\simeq k,\\
Y_\pm(\tilde{\boldsymbol{q}})\simeq -t_M[1+\frac{1}{3}\tilde{\boldsymbol{q}}^2\pm \frac{1}{9}  \tilde{q}_y(3 \tilde{q}_x^2- \tilde{q}_y^2)], \\
Z_\pm(\tilde{\boldsymbol{q}})\simeq t_M^2[\tilde{\boldsymbol{q}}^2\pm \frac{2}{3} \tilde{q}_y(3 \tilde{q}_x^2- \tilde{q}_y^2)].
\end{array}
\end{equation}
 such that we can rewrite the electron and hole bands
 \begin{equation}
\begin{array}{l}
E_1(\tilde{\boldsymbol{q}})=k+\frac{t_M}{2}[1+\frac{1}{3}\tilde{\boldsymbol{q}}^2+ \frac{1}{9}  \tilde{q}_y(3 \tilde{q}_x^2- \tilde{q}_y^2)+\sqrt{\tilde{\boldsymbol{q}}^2+ \frac{2}{3} \tilde{q}_y(3 \tilde{q}_x^2- \tilde{q}_y^2)}],\\
E_2(\tilde{\boldsymbol{q}})=k+\frac{t_M}{2}[1+\frac{1}{3}\tilde{\boldsymbol{q}}^2+ \frac{1}{9}  \tilde{q}_y(3 \tilde{q}_x^2- \tilde{q}_y^2)-\sqrt{\tilde{\boldsymbol{q}}^2+ \frac{2}{3} \tilde{q}_y(3 \tilde{q}_x^2- \tilde{q}_y^2)}],\\
E_3(\tilde{\boldsymbol{q}})=k-t_M[1+\frac{1}{3}\tilde{\boldsymbol{q}}^2+ \frac{1}{9}  \tilde{q}_y(3 \tilde{q}_x^2- \tilde{q}_y^2)].
 \end{array}
\end{equation}
Similarily for the hole bands $H_1>H_2>H_3$ one obtains,
\begin{equation}
\begin{array}{l}
H_1(\tilde{\boldsymbol{q}})=-k-t_M[1+\frac{1}{3}\tilde{\boldsymbol{q}}^2- \frac{1}{9}  \tilde{q}_y(3 \tilde{q}_x^2- \tilde{q}_y^2)],\\
H_2(\tilde{\boldsymbol{q}})=-k+\frac{t_M}{2}[1+\frac{1}{3}\tilde{\boldsymbol{q}}^2- \frac{1}{9}  \tilde{q}_y(3 \tilde{q}_x^2- \tilde{q}_y^2)-\sqrt{\tilde{\boldsymbol{q}}^2- \frac{2}{3} \tilde{q}_y(3 \tilde{q}_x^2- \tilde{q}_y^2)}],\\
H_3(\tilde{\boldsymbol{q}})=-k+\frac{t_M}{2}[1+\frac{1}{3}\tilde{\boldsymbol{q}}^2- \frac{1}{9}  \tilde{q}_y(3 \tilde{q}_x^2- \tilde{q}_y^2)+\sqrt{\tilde{\boldsymbol{q}}^2- \frac{2}{3} \tilde{q}_y(3 \tilde{q}_x^2- \tilde{q}_y^2)}].
 \end{array}
\end{equation}

The Dirac cones at $\kappa_S$ (between $E_1$ and $E_2$ and betwen $H_2$ and $H_3$) have a dispersion relation $\pm |\tilde{\boldsymbol{q}}|$, i.e. they are isotropic and have a velocity $v_{\kappa_S}=1/|t_M|$.

Expanding $E_2({\bm q})$ to quadratic order around the saddle point at $\boldsymbol{q}^*=(0,0.78 t_M)$, one finds
\beq
E_2({\bm q}^*+{\bm q})\simeq E_2^* +\alpha_x q_x^2 + \alpha_y q_y^2 ,
\eeq
with $E_2^*=k+0.305 t_M$, $\alpha_x=- 0.888/t_M $ and $\alpha_y= 0.342/t_M$. 
From these expressions, one deduces that the saddle point energy is shifted by an amount $-0.195 t_M$ from the Dirac point at $\kappa_S$ with energy $E_2(\kappa_S)=k+t_M/2$. This shift is much larger than that of the saddle point in $E_1$ from the Dirac points at $m_S$. The saddle point  on the $E-2$ band is also much less anisotropic than the saddle point in $E_1$ (there is a factor 2.6 between the two curvatures instead of $1/(2 t_M)^2\sim 100$).

Expanding $H_2({\bm q})$ to quadratic order around the saddle point at $\boldsymbol{q}^*=(0,-0.78 t_M)$, one finds that  the curvatures $\alpha_x$ and $\alpha_y$ are the same as for $E_2$. 
%and only the saddle point energy $H_2^*=-k+0.305 t_M$ is different. 

\subsection{Effective parameters}
We now discuss the effective parameters of the Dirac cones at $m_S$ (velocities $c_x$, $c_y$ and geometrical average $v_S\equiv \sqrt{c_x c_y}$ in units of $v$) and of the saddle points (curvatures $\alpha_x=1/(2 m_{xx})$, $\alpha_y=1/(2 m_{yy})$ and geometrical average $\alpha\equiv \sqrt{|\alpha_x \alpha_y|}$ in units of $v/G$) in $E_1/H_1$ close to $m_S$ and in $E_2/H_2$ close to $\kappa_S$. We fit these parameters on the numerically obtained band structure (see Table~\ref{tab}). The validity of the quadratic expansion near a saddle point is up to an energy $\varepsilon_c\sim 10^{-2} vG$.
\begin{table}[h!]
\center
\begin{tabular}{ c || c | c | c | c | c | c | c | c } 
 $M_B$& $c_x$ & $c_y$ & $v_S$ & $\alpha_x$ & $\alpha_y$ & $\alpha$  & $\alpha v_S^2$ & $|t_M| \alpha/v_S^2$  \\ \hline \hline
$E_1$ & 1 & 0.1 & 0.32  & -70 & 1 & 8.4 & 0.84 & 5.4\\ \hline   
$E_2$ & $''$ & $''$ &  $''$ & 16 & -5 & 8.9 & 0.89 & 5.7\\ \hline  
$H_1$ & 1 & 0.15 & 0.38 & 60 & -1 & 7.7  & 1.16 & 3.3\\ \hline  
$H_2$ & $''$ & $''$ & $''$  & 12 & -6 & 8.5   &1.3 & 3.6 \\ \hline \hline
pred. $E_1/H_1$ & 1 & 0.13  & 0.36 & $\mp 61$ & $\pm1$ & 7.8 & 1 & 3.9 \\  \hline
pred. $E_2/H_2$ & $''$ & $''$ &  $''$ & 14 & -5 & 8.6 & 1.1 & 4.3
%prediction $H_1$ & $''$ & $''$ &  $''$  & 61 & -1 & 7.8 & 1 & 3.9
\end{tabular}
\quad
\begin{tabular}{ c || c | c | c | c | c | c | c  | c} 
 $M_A$& $c_x$ & $c_y$ & $v_s$ & $\alpha_x$ & $\alpha_y$ & $\alpha$ &$\alpha v_S^2$ & $|t_M| \alpha/v_S^2$  \\ \hline \hline
$E_1$ & 1 & 0.07 & 0.26 & -150 & 1 & 12  & 0.85 & 6.1  \\ \hline   
$E_2$ & $''$ & $''$ & $''$ & 25 & -7 & 13.2 & 0.92 & 6.6 \\ \hline  
$H_1$ & 1 & 0.07 & 0.26 & 200 & -1 & 14  & 0.98 & 7 \\ \hline  
$H_2$ & $''$ & $''$ & $''$ & 23 & -10 & 15  & 1.1  & 7.6 \\ \hline \hline
pred. $E_1/H_1$ & 1 & 0.07  & 0.26 & $\mp204$ & $\pm1$ & 14.3 & 1 & 7.1 \\  \hline
pred. $E_2/H_2$ &  $''$ & $''$ &  $''$& 25 & -10 & 15.8 & 1.1 & 8.2 
%prediction $H_1$ &  $''$ & $''$ &  $''$& 204 & -1 & 14.3 & 1 & 7.1 
\end{tabular}
\caption{Effective parameters for sample $M_B$ with $t_M=-0.064$ (left) and for sample $M_A$ with $t_M=-0.035$ (right). Velocities are in units of $v$ and curvatures in units of $v/G$.}
\label{tab}
\end{table} 

This agrees well with the  analytical predictions for the anisotropic velocities $c_x=1$ and $c_y=2|t_M|$ of the Dirac points at $m_S$ and for the curvatures of the saddle points in the bands $E_1/H_1$ (close to $m_S$), $\alpha_x=\mp 1/(2t_M)^2$ and $\alpha_y=\pm1$ (i.e. $\alpha=0.5/|t_M|$) and in the bands $E_2/H_2$ (close to $\kappa_S$), $\alpha_x=-0.888/t_M$ and $\alpha_y=0.342/t_M$ (i.e. $\alpha\simeq 0.55 /|t_M|$). 
It means that $v_S^2=2|t_M|$, $\alpha\simeq 0.5/|t_M|$ so that $\alpha v_S^2\simeq 1$. 
From the two energies $t_M$ and $v_S^2/\alpha = 2 m_{eff} v_S^2$, one can form the dimensionless ratio
\beq
|t_M|\frac{\alpha}{v_S^2}\simeq \frac{1}{4|t_M|} \, ,
\eeq
which is $\sim 4$ for $M_B$ and $\sim 7.5$ for $M_A$. We use this estimate $|t_M|\frac{\alpha}{v_S^2}\sim 4 (M_B) - 7.5 (M_A)$ repeatedly in the following.

\subsection{Density of states and susceptibility singularities at a saddle point}

In the following we present analytical results on  the paramagnetic singularity of the susceptibility  at the two types of saddle points  encountered in the moir\'e  band structure of graphene discussed above. The first one is the ''ordinary'' $A_1$ saddle point, where energy exhibits a maximum along one axis and a minimum along the perpendicular one. The second one is the $C_3$ saddle point, where the band curvature is zero in all directions. This point is surrounded by three maxima separated by three valleys at $120^{\circ}$  angles. 
\medskip

For clarity, we compute in parallel both density of states and susceptibility, whose expressions bear similarities~:
\begin{eqnarray} \rho &=& \int  \delta(\epsilon(\bs p)-\epsilon)\, {d^2p \over 4 \pi^2} \\
 \chi &=& \int   \delta(\epsilon(\bs p)-\epsilon)  \chi(p_x,p_y)\,   {d^2p \over 4 \pi^2}   \end{eqnarray}
with 
 \be \chi(p_x,p_y)= {\mu_0 e^2 \over  12} \left({\partial ^2\epsilon \over\partial p_x^2}{ \partial ^2\epsilon  \over \partial p_y^2}- ({\partial ^2\epsilon  \over \partial p_x \partial p_y})^2\right) \ee
$\chi$ is the so-called Landau-Peierls contribution to the susceptibility, neglecting interband effects \cite{Peierls}.

\subsubsection{DOS and susceptibility singularities at a $A_1$ saddle point}

We  consider the simple saddle point in the dispersion relation modeled as :
\be \ep(p_x,p_y)=\alpha_x p_x^2 - \alpha_y p_y^2\ . \label{E4} \ee
$\alpha$ has the  dimension of an inverse mass $\alpha=1/(2m)$. For generality,
we consider the anisotropic case relevant for our situation and choose $0 < \alpha_y < \alpha_x$ and  we introduce a momentum cutoff $p_c$. The integrated density of states (DOS)  is given by~:
\be N(\ep)={1 \over 4 \pi^2}  \int dp_x dp_y \ee
with the constraint
\begin{eqnarray}
 & 0 <\ep(p_x,p_y) < \ep  \qquad & \qquad \text{for} \  \ep >0     \nonumber \\
 & \ep <\ep(p_x,p_y) < 0  \qquad & \qquad \text{for} \ \ep <0     \label{region11}
\end{eqnarray}
 
After momentum integration and derivation with respect to the energy, the density of states is given by
\be   \rho(\ep)=  {1 \over 2 \pi^2 \sqrt{\alpha_x \alpha_y}} \ln {\sqrt{\alpha_y} p_c + \sqrt{\alpha_y p_c^2 + \ep} \over \sqrt{|\ep|}} \ . \label{rhomoins1}\ee
At energy well below the cutoff $\sqrt{\alpha_y} p_c$, it takes the simple form

\be   \rho(\ep)=   {1 \over 4 \pi^2 \sqrt{\alpha_x \alpha_y}} \ln {4 \alpha_y p_c^2 \over |\ep|} \ .    \label{rhomoins2}\ee

Within the Landau-Peierls approximation, the susceptibility is obtained from a similar  calculation with an extra factor $\chi(p_x,p_y)=  (\mu_0 e^2 /3) 
 \alpha_x \alpha_y$, leading to:

\be   \chi(\ep)={\mu_0 e^2  \over 12 \pi^2}  \sqrt{\alpha_y \alpha_x}   \ln {4 \alpha_y p_c^2 \over |\ep|}\label{chimoins}\ee

\subsubsection{DOS and susceptibility singularities at a $C_3$ saddle point}

The energy varies as a cubic function of $p_x$ and $p_y$, which is convenient to parametrize in polar coordinates as: 
\be  \ep(p_x,p_y)=\beta \ (p_x^3 - 3 p_x p_y^2) \equiv \beta p^3 \cos 3 \theta \label{E6}\ee
with an energy  cutoff  $\ep_c= \beta p_c^3$.   With integration regions constrained by eq. \ref{region11}, the integrated DOS  is given by

\be N(\ep)= {6 \over 4\pi^2} \int_{\beta p^3 \cos 3 \theta < \ep } d^2p = 
{ 3 \over 4 \pi^2}\, \left({\ep\over \beta}\right)^{2/3} \  \int_0^{\pi/6} {d \theta \over( \cos 3 \theta )^{2/3}}   \label{NC3-1}\ee
The integral converges and yields: $\sqrt{\pi}\Gamma(7/6) /\Gamma(2/3)=1.214$, so that we get finally for the DOS :

\be {\rho(\ep)= {C \over \beta^{2/3} |\ep|^{1/3}}}\ee
with $C= 0.0615$.
 
\bigskip

For the integrated susceptibility $X(\ep)$, the integrand  has a multiplicative term in $3 \mu_0 e^2 \beta^2  p^2$. This yields in polar coordinates:
\be X(\ep)= {9 \over 2 \pi^2}  \mu_0 é^2 \beta^2 \int_0^{\theta_c} d \theta \int_{\beta p^3 \cos 3 \theta < \ep} p^3 \, dp \ .  \ee
Integrating over  $p$ with the condition $\beta p^3 \cos 3 \theta < \ep$, leads to:

\be X(\ep)=  {9 \over 8 \pi^2} \mu_0 e^2 \beta^{3/2} \ep^{4/3} \int_0^{\theta_c} {d \theta \over (\cos 3 \theta)^{4/3} } \ee
with the cutoff $\theta_c \simeq \pi/6 - \ep/(3 \ep_c)$.
%Je trouve que (pas sur de la valeur exacte du 3/4, mais c'est excellent num\'eriquement):
We use the expansion~:

 \be \int_0^{\pi/6 - x} {d \theta \over (\cos 3 \theta)^{4/3}} \quad \limite{x \rightarrow 0} \quad  {1 \over (3 x)^{1/3}} - b \ .  \ee
with $b= \sqrt{\pi} \Gamma(5/6)/\Gamma(1/3) \simeq 0.747$ and we find 
\be X(\ep)=  {9 \mu_0 e^2 \over 8 \pi^2}  \, \beta^{2/3} \ep^{4/3} \left( {\ep_c^{1/3} \over \ep^{1/3}} - b\right) \ee
The susceptibility at energy $\ep$ is obtained by derivation and reads:
\be { \chi(\ep)= {9 \over 8 \pi^2} \mu_0 e^2 \,   \beta^{2/3}\,(\ep_c^{1/3}  - c|\ep|^{1/3} )} \label{eq:chi6}  \ee
with $c=4 b /3 \simeq 1$.
The susceptibility is  therefore finite at its maximum and depends on the cut-off $\ep_c$.

\subsubsection{ Gaussian broadening }
The effect of disorder can be accounted for by a gaussian distribution of chemical potentials of width $\sigma$. The susceptibility becomes
\be \chi_\sigma(\mu)= \int \chi (\mu') \ P(\mu - \mu') d\mu' \ee
with
\be P(\mu)={e^{-\mu^2 \over 2 \sigma^2} \over \sqrt{2 \pi} \sigma }  \label{eq:gauss} \ee

The logarithmic singularity in eqs.(\ref{rhomoins2},\ref{chimoins}) is smoothed  according to:
\be \ln |\ep|  \longrightarrow    \int \ln |\mu'| {e^{-(\mu-\mu')^2 \over 2 \sigma^2} \over \sqrt{2 \pi} \sigma } d\mu' \ee
%L'int\'egrale n'est pas simple. 
For $\mu=0$,   the divergence    $1/\ln |\ep|$   is cut as
 $1/\ln A \sigma$  
with $A= \sqrt{{1 \over 2}} e^{-\gamma/2} \simeq 0.53 $ and  $\gamma =0.577...$ is the Euler constant. 

\medskip

 Similar calculation for the $C_3$ singularity yields:
\be |\ep|^{1/3} \longrightarrow  \int  |\mu'|^{1/3}  {e^{-(\mu-\mu')^2 \over 2 \sigma^2} \over \sqrt{2 \pi} \sigma } d\mu' \ee

    For $\mu=0$, %in units of $\ep_c$, 
		one finds that~:
	
	\be { \chi(\mu=0)= {9 \over 8 \pi^2} \mu_0 e^2 \,   \beta^{2/3}\,(\ep_c^{1/3}  - d\sigma^{1/3} )}  \label{chimu0C6}\ee

with $d= c \, 2^{1/6} \Gamma(2/3) /\sqrt{\pi}\simeq 0.854$.

\subsection{Orbital susceptibility and comparison with experimental data}

One important parameter is the  energy width $\sigma_s$ of the satellite Dirac peak which limits the amplitude of the susceptibility peaks.
In Fig.\ref{fig:bonus}  in section II.C, we show how to extract $\sigma_s$ from the  gate voltage  derivative of the magnetisation  $ M'(V_g) \propto\mu M'(\mu)$.    The spacing  between the  maximum and the minimum   of  $M'(\mu)$  fitted by the derivative of a gaussian centered at $\mu_S$ around $\mu $ yields $2\sigma_S = 9 \pm 1$ meV  for both samples.
In the following  for quantitative comparisons we mostly focus on the  sample $M_B$, with the largest moiré potential period and amplitude, for  which the data is the most reliable due to the larger energy spacing between paramagnetic and diamagnetic peaks.

\subsubsection{Secondary Dirac peak versus main Dirac peak}
The ratio of susceptibility peaks at  the sDP  $\chi_S$ and the main Dirac peak $\chi_M$ reads: 
\beq
r\equiv \frac{\chi_S}{\chi_M}=\frac{3 v_S^2/\sigma_S}{v^2/\sigma_0}=3\times 2 |t_M|\times \frac{\sigma_0}{\sigma_S} = 3\times 2(0.035-0.064)\times 1.5\approx 0.32 (M_A) -0.58 (M_B)
\eeq
%or of the derivatives
%\beq
%r'\equiv \frac{\chi_S '}{\chi_D '}=\frac{3 v_S^2/\sigma_S^2}{v^2/\sigma_0^2}=3\times 2 |t_M|\times \frac{\sigma_0^2}{\sigma_S^2} = 3\times 2(0.035-0.064)\times 1.5^2\approx 0.47 (M_A) -0.86 (M_B)
%\eeq
Experimentally, $\sigma_0=80$~K $=6.9$~meV, and $\sigma_S=4.5$~meV so that $\sigma_S/|t_M|\sim 0.2 (M_B) - 0.3 (M_A)$. 

The measured ratio is $r=0.33\pm 0.1$,  for $M_B$ which is of the order but  1.7 times lower than  the expected   value given above,  we show in the next section that this disagreement is probably due to an underestimation of  the diamagnetic peak $\chi_S$  very  close to the inner Vignale  paramagnetic peak (on the low doping side). %a bit far (factor 3.6 for $M_A$ and 1.8 for $M_B$). 
%If we are able to estimate (separately) $r'$ and $r$ then we get an estimate of 
%\beq
%\frac{\sigma_0}{\sigma_S}= \frac{r'}{r} .
%\eeq
%If $\sigma_0 = \sigma_S$, then $r=r'=6|u_0|=6(0.035-0.064)\approx 0.21 (M_A) -0.38 (M_B)$.
The same calculation gives $r=$ $0.32$ for $M_A$ which is also larger than the measured value  of the order of 0.1 with  a very large error bar (see tables in section II).%In principle one could  perform a similar analysis on sample $M_A$, however the experimental error bar on this sample  is very large i.e. of the order of the peak amplitude, see tables in section II.

%The measured ratio is $r= 0.33$, which is a bit far (factor 3.6 for $M_A$ and 1.8 for $M_B$). 
%If we are able to estimate (separately) $r'$ and $r$ then we get an estimate of 
%\beq
%\frac{\sigma_0}{\sigma_S}= \frac{r'}{r} .
%\eeq
%If $\sigma_0 = \sigma_S$, then $r=r'=6|u_0|=6(0.035-0.064)\approx 0.21 (M_A) -0.38 (M_B)$.

\subsubsection{Vignale paramagnetism versus McClure diamagnetism}
The Vignale paramagnetic susceptibility $\chi_V \sim \alpha \ln (\epsilon_c/\sigma_s)$ and the McClure diamagnetic susceptibility (at the sDP) $\chi_S \sim - v_S^2/\sigma_s$, where $\sigma_s$ is the disorder broadening. Their ratio (up to numerical factors) is 
\beq
\frac{\chi_V}{|\chi_S|}\sim \frac{\alpha}{v_S^2} \sigma_s  \ln (\epsilon_c/\sigma_s)\sim  |t_M|\frac{\alpha}{v_S^2} \times \frac{\sigma_s}{|t_M|} \approx (4-7.5)\times (0.1-0.15)\approx 0.4 (M_B)-1.1 (M_A) .
\eeq 
This is just an order of magnitude estimate. However, for the precise ratio between the maximas of the derivative $\chi'$, we find:
\beq
r_{VD}=\frac{\chi_V '}{|\chi_S '|}\approx \frac{6}{3}\times \frac{ 0.0065 \alpha/\sigma_s}{0.013 v_S^2 /\sigma_s^2} \approx \frac{\alpha}{v_S^2} \sigma_s = |t_M| \frac{\alpha}{v_S^2} \times \frac{\sigma_s}{|t_M|} \approx 0.4 (M_B)-1.1 (M_A) ,
\eeq
where $6/3$ accounts for the 6 saddle points versus the 3 sDPs and $\sigma_s/|t_M|\sim 0.1$. The relative magnitude of the orbital paramagnetism and diamagnetism for the moir\'e band structure is therefore given by the following important dimensionless ratio
\beq
\frac{\alpha}{v_S^2} \sigma_s = \frac{vG}{4 t_M} \frac{\sigma_s}{t_M}, 
\eeq
It is interesting to note that $r_{VD}$ decreases with $  1/G \propto a_M$ and $t_M$. This is why it is expected to be larger for the $M_A$ sample  than$M_B$ in agreement with our experimental results.  In particular  all paramagnetic peaks on bands $E_1$ , $H_2$  and $E_2$  are visible for $M_A$.
 A different behavior is observed for the $M_B$ sample:  saddle points on the $E_1$ and $H_1$ bands are nearly not detectable whereas  instead paramagnetic peaks are clearly observed on 
the $E_2$,$H_2$ bands and the ratio $r _{VD} = 0.8 \pm 0.2 $,  $0.35 \pm 0.15$  can be estimated,  see section II below. A similar  quantitative estimation is delicate on   $M_A$ :   on the  one hand the amplitude of the diamagnetic peak at sDPs is  difficult  to estimate and on the other hand  the large doping region  of saddle points in the $E_2$ and $H_2$  bands correspond to large gate voltages of the order of $\pm 20V$ for which  the insulating properties of the hBN top  layer   become not reliable and therefore experimental data are not reproducible. This explains the large error bars in Table II of section II.

%This is because $\alpha/v_S^2$ is actually of order $t_M a_M^2/(v t_M a_M)\sim (c_x/c_y)/t_M \sim 10/t_M$ and not  $\alpha/v_S^2\sim t_M a_M^2/(t_M a_M)^2 =1/t_M$ (as previously assumed). 
%Including disorder broadening $\sigma$, the ratio $\chi_\text{V}/|\chi_\text{M}|$ is proportional to $\alpha \ln( \epsilon_c/\sigma)/(v_S^2/\sigma) \sim 6.5 \sigma/ t_M\sim 1$ 
%Remark (for us): Compared to the initial estimation of $\chi_\text{V}/|\chi_\text{M}|$, the changes are
%\beq
%\frac{1}{4 \pi^2} \times 2 \times \frac{6}{3} \times (4-7.5) \approx 0.4 (M_B)-0.8 (M_A),
%\eeq
%where $1/(4\pi^2)$ is a typo in the computation of the Vignale susceptibility, $2$ is a typo in the McClure susceptibility, $6/3$ is the ratio between the number of saddle points and that of Dirac points, and $4-7.5$ is a factor due to the effective velocities and curvatures.

\subsection{Vignale orbital paramagnetism versus Pauli spin paramagnetism at the saddle points}
Vignale's susceptibility is $\chi_V = \frac{\alpha^2}{3}\rho(\epsilon)$ where $\rho$ is the DoS and we have taken $\mu_0\equiv 1$ and $e\equiv 1$. Pauli's susceptibility is $\chi_P=\frac{g}{2}\mu_B^2 \rho(\epsilon)$ with the Bohr magneton $\mu_B = \frac{e \hbar}{2 m_e} = \frac{1}{2 m_e}$, where $m_e$ is the bare electron mass and $g\approx 2$ for graphene. Their ratio is:
\beq
\frac{\chi_V}{\chi_P}=\frac{\alpha^2/3}{\mu_B^2}=\frac{4}{3}\alpha^2 m_e^2=\frac{1}{3}\left(\frac{m_e}{m_{eff}} \right)^2=\frac{1}{3u_0^2} \left(\frac{m_e v}{\hbar G}\right)^2\approx 14000 (M_B)-27000 (M_A) \sim 10^4 .
\eeq
In other words, since a typical effective mass at the saddle point of the moir\'e band structure  is $m_{eff}=1/(2\alpha) \sim 4.10^{-3} m_e$  (i.e. $m_e/m_{eff}\sim 200$), the spin paramagnetism is completely negligible compared to the orbital paramagnetism. 
%The anisotropy is also quite large as $m_x\sim 4.10^{-4} m_e$ and $m_y\sim 4.10^{-2} m_e$, i.e. $m_y/m_x\sim 100$.
We note that $\frac{1}{3}\left(\frac{m_e}{m_{eff}} \right)^2$ is the famous ratio between orbital and spin magnetic susceptibility usually discussed near a band edge when comparing Landau diamagnetism and Pauli paramagnetism.

%%%%%%%%%%%%%%%%%%%%%%%%%%%%%%%%%%%%%%%%%%

%\title{VH singularities, SM}

%\section*{Supplementary }

\section{More details on  experimental results.}

\subsection{Fabrication and characterization of the investigated samples.}
\begin{figure}[h!]
\begin{center}
\includegraphics[width=8cm]{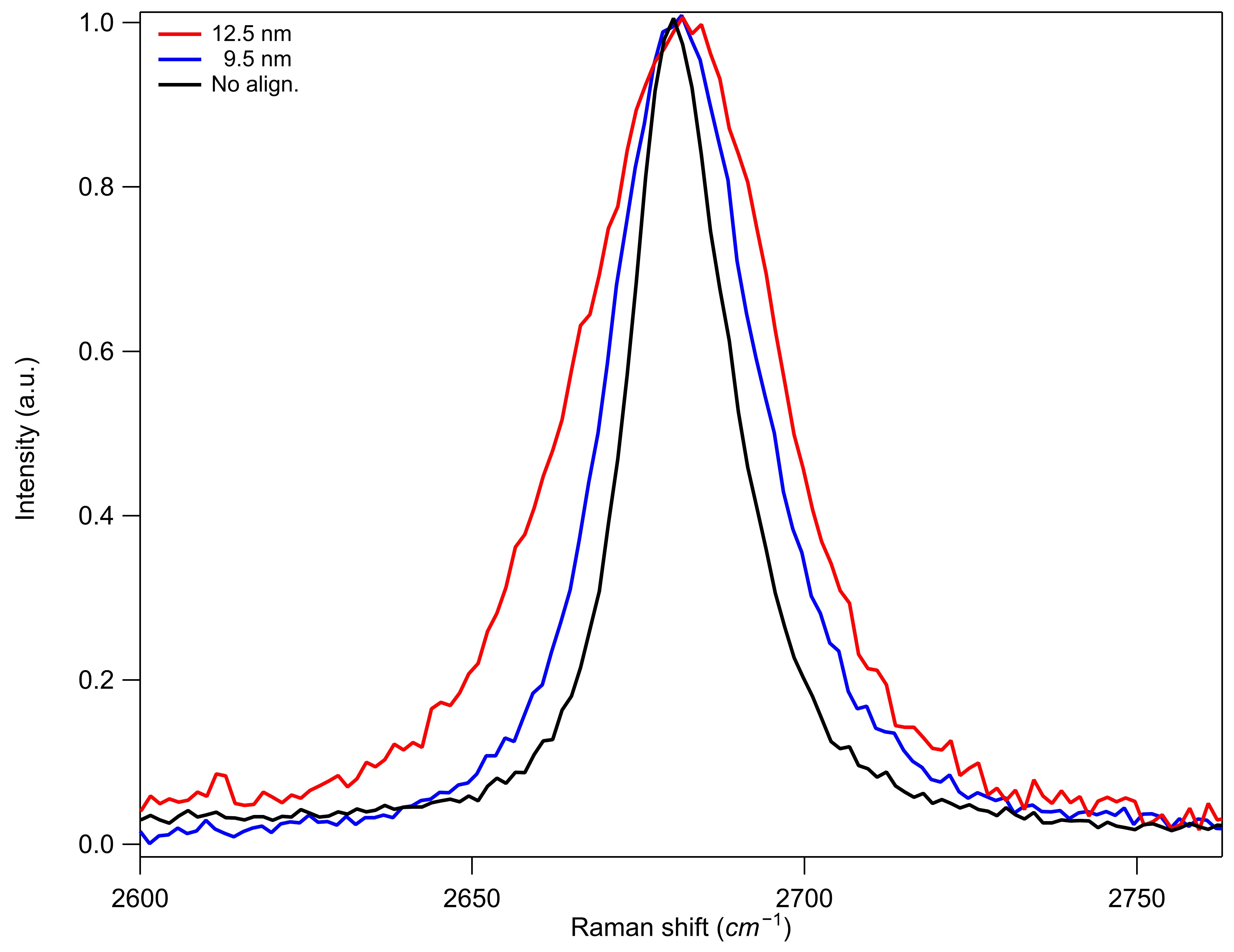}\\
\caption{ Signatures of the moiré potential  on the width of the Raman 2D peak. } 
\label{Raman}
\end{center}
\end{figure}
We start from a selected exfoliated graphene monolayer with a hBN flake. 
We first identified  long straight edges in both graphene and hBN layers. These edges  follow the  crystallographic axes of each  honey-comb lattice. Therefore, by aligning  these straight edges, one has equal probability %we have 50$\%$ chance 
that graphene and hBN are aligned  %at 0$^{\circ}$ and 50$\%$ chance 
or that their respective alignment form a 30$^{\circ}$ angle.  In order to guarantee the alignment, two different methods were followed for the two samples $M_A$ and $M_B$ investigated. 

For sample $M_B$, a flake of graphene with a long straight edge of about 45$\mu m$ long was cut in three parts of roughly 20, 5 and 20$\mu m$ respectively. The first part was aligned  and picked-up with a straight edge of a big flake of hBN. Then covered by a bottom misaligned hBN. The second and third parts were rotated  with an angle of 30$^{\circ}$ with respect to the hBN edge,  picked up with another part of the same top hBN flake and  finally dropped onto a misaligned hBN. Raman spectroscopy measurements allowed us to determine which of the two samples  is the one with the largest moiré constant by measuring the  width of the 2D peak. This sample was selected and  deposited on the top of the GMRs based  magnetisation detector,  using the standard dry transfer techniques.

In the case of sample $M_A$, a large  hBN flake was cut into two parts along one  main crystallographic edge (this was done by opening  a narrow slit though the hBN flake using  electron-beam lithography followed by reactive ion etching ) %was used to open a window in a resist (PMMA) and latter, the exposed BN was removed
 A single graphene flake was then aligned along the straight edge of one of the halves of the hBN flake,   and encapsulated beween this part and the other half previously rotated by $30^{\circ}$. The presence of a long-range moiré pattern, was confirmed by Raman spectroscopy associated this changes with 
For both samples  Raman  experiments enabled us to determine the  moiré lattice parameters. According to \cite{Ribeiro} (see also \cite{Eckermann}),  the full width at half maximum (FWHM) of the 2D peak of the Raman spectrum of graphene is very sensitive to  the folding of the phonon structure due to the moiré pattern. This creates copies of the 2D peaks with small differences in their Raman shift. This is reflected by an increase of the FWHM of the 2D peak which varies linearly with the  superlattice period, $a_M$ leading to the relation FWHM$_{2D}=2.7a_M+0.77$.
 % In the case of the samples investigated here, Raman measurements revealed FWHM$_{2DM_A}=26.5\ cm^{-1}$ and FWHM$_{2DM_B}=34.9\ cm^{-1}$ as shown in the SM. This corresponds to $a_{M_A}=9.5nm\pm0.5nm$ and $a_{M_B}=12.5nm\pm0.5nm$.

 %This moiré lattice parameter $a_M=(1+\delta)a_G/(\sqrt{2(1+\delta)[1-\cos{\theta}]+\delta^2})$,   depends on the twist angle $\theta$ between the hexagonal lattices of graphene and hBN ( $a_G =0.25$ nm is the lattice constant of graphene and $\delta=0.017$ is the ratio between graphene and hBN lattice constants). 

From Fig. \ref{Raman},  we found the FWHM  of the D peak to be 26.5$cm^{-1}$ for M$_A$ sample and 34.9$cm^{-1}$ for M$_B$.  
From these values, using the relation given  above relating the width of the Raman D peak to the moiré period, we can deduce  for the 2 samples A and B,  $a_M =  9.5 nm$  and $ 12.5 nm $ nm  as well the twist angle $\theta$ between the hexagonal lattices of graphene and hBN  acording to   the relation $a_M=(1+\delta)a_G/(\sqrt{2(1+\delta)[1-\cos{\theta}]+\delta^2})$,   where $\delta=0.017$ is the ratio between graphene and hBN lattice constants). We find $\theta_A = 1.1$° and $\theta_B = 0.6 $°.

The period of the moiré lattices coincides with those expected from the magnetisation curves. As shown in Fig.3 of the main paper, the diamagnetic peaks are located very close to the doping (or $V_g$) at which the density is equal to 4 electrons (or holes) per moiré-unit cell. In contrast, transport measurements show a satellite peak at a smaller density. Since the geometry of the two experiments differ, we may be seeing a slightly different effective moiré period(of about 0.5nm) in each experiment. This difference might be caused by local strain. In any case, the difference in the position in doping between magnetisation and resistance measurements can be explained by considering the error bars of the calibration of the moiré-length/FWHM relation in \cite{Ribeiro}.

\subsection{Effect of the gate voltage modulation}

We first present  in Fig.7 the  gate voltage derivative of the resistance of sample $M_B$. This allows us to directly compare the positions (in gate voltage) of the main and secondary Dirac peaks with the corresponding diamagnetic peaks. We observe antisymmetric peaks both in the resistance and the magnetisation at the Dirac peak. At higher doping, we see the derivative of the secondary Dirac peaks in the resistance . In magnetization, we observe both seconday McClure peaks and paramagnetic peaks. When this magnetization curve is integrated, it gives the curves shown in Fig.3 of the main text for different values of magnetic field. From this data  taken at different values of magnetic field between $\pm 0.1$ and $\pm 0.2$ T one  can appreciate the reproducibility of the intensity and positions of the  diamagnetic secondary  McClure  peak and the outer paramagnetic peaks, whereas the inner paramagnetic peaks are nearly invisible at low field.

\begin{figure}
    \centering
    \includegraphics[width=0.5\linewidth]{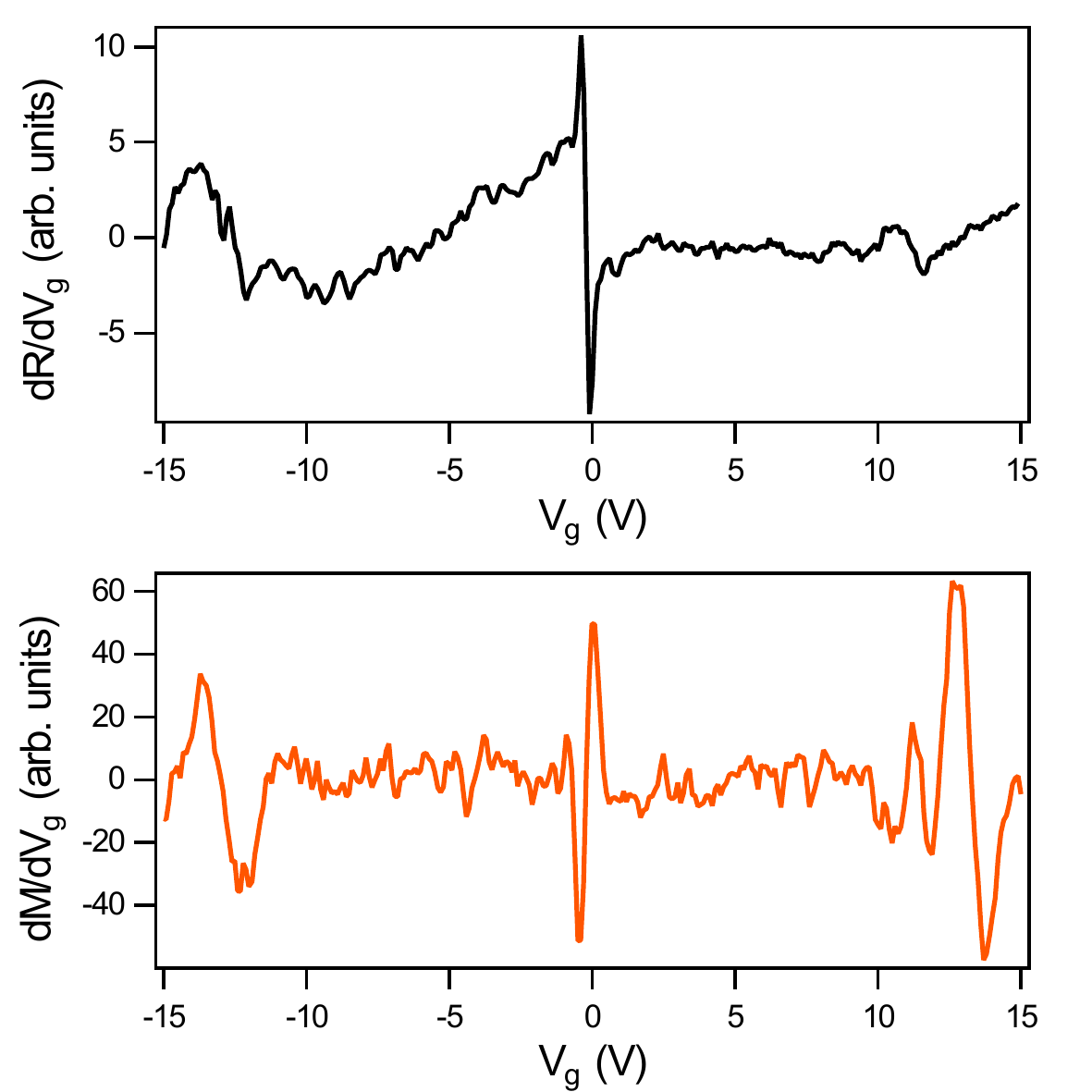}
    \caption{Top: derivative of the resistance as a function of the gate voltage for sample $M_B$. Bottom: derivative of the magnetization as a function of the gate voltage for the same sample.}
    \label{fig:dRdM}
\end{figure}
\begin{figure}
    \centering
    \includegraphics[width=0.5\linewidth]{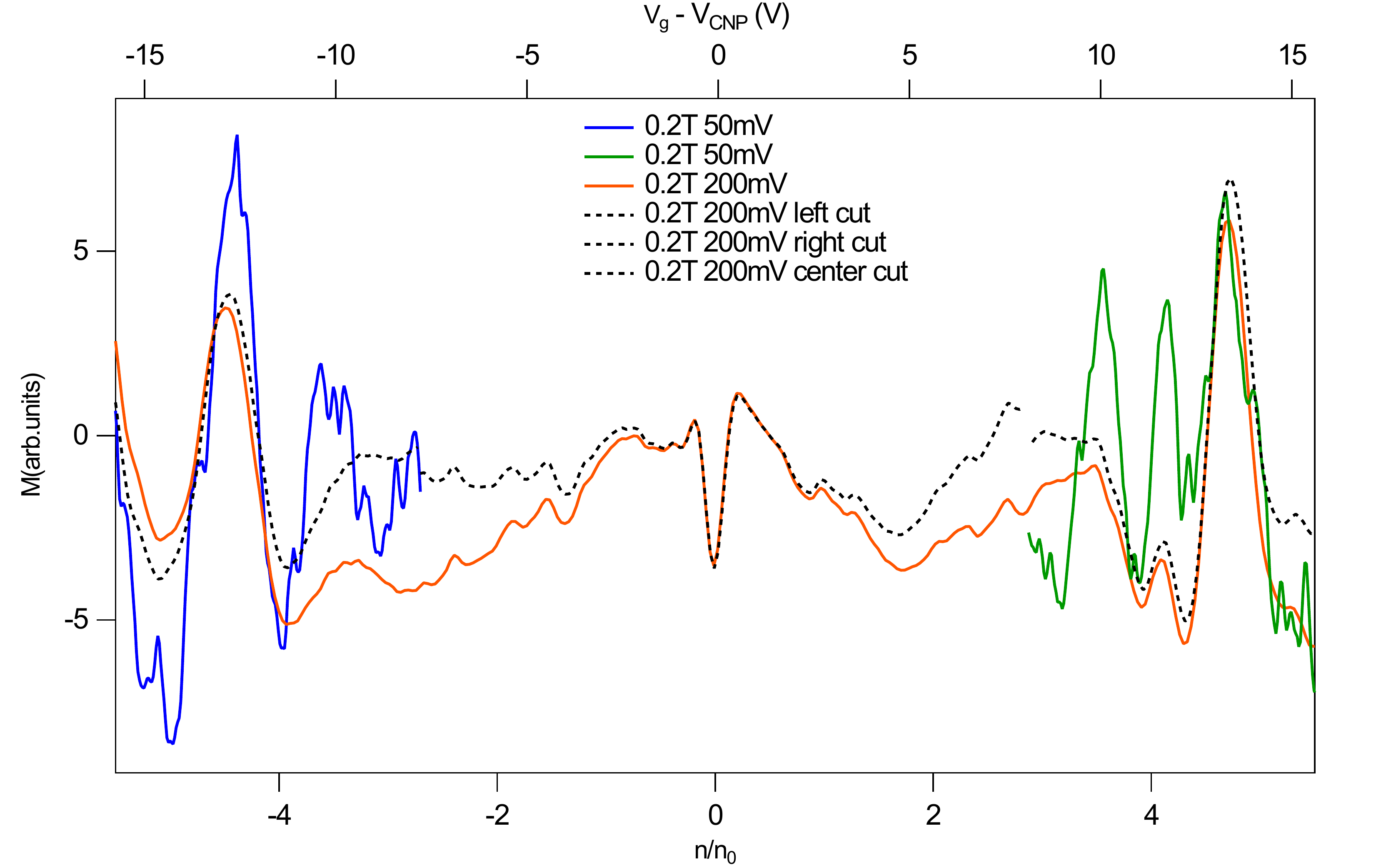}
    \caption{Comparison of the magnetization of sample $M_B$, with  different gate voltage  modulations, for an external field of $0.2$ T. In orange, the curve for $200$  millivolt integrated in the full range of gate voltage investigated ($-15.5$ to $15.5$  volt), see Fig.3 in the main text. In blue, the electron's (left) sDP region measured at $50$ millivolt and integrated in a range between $-16$ to $-8$  volt. In green, the hole's (right) sDP region measured with  $50$ millivolt gate voltage modulation and integrated in a range between $8$ to $16$ volt. In dashed black, the integration of the data obtained with $200$ millivolt gate voltage modulation was split into 3 regions: left ($-16$ ; $-8$) volt, center ($-8$ ; $8$) volt and right ($8$ ; $16$) volt.  }
    \label{fig:my_label}
\end{figure}

We now consider on the same data  the effects of modulation and integration. Figure \ref{fig:my_label} compares the integrated curve of data in \ref{fig:dRdM}  after subtraction of a linear background, with data taken obtained for a smaller range of gate voltage and with a smaller modulation. We can notice that the higher modulation allows us to obtain a smoother curve with less noise, but certain structures get rounded in a non-negligible way.  We also investigated  the effect of the range of  gate voltage along which  integration is performed.  We compare the curves obtained after integrating the data  in the whole range of $V_g$  investigated (between -15 to 15V) with the curve obtained after  splitting  the data in 3 pieces along the $V_g$ axis.  Whereas the  main magnetization peaks are unchanged, we find that  integrating over the full range of gate voltage generates an extra negative contribution  on the magnetisation which is not seen when the integration range is reduced to 10V.

\subsection{Additional data for different values of magnetic field}
We present here  magnetisation data not shown in the main part of the paper. Magnetisation is expressed in units of the magnetic field detected on the GMR sensor.
For the measurements  on $M_B$ the experimental conditions (dc current though the GMR detector as well as gate voltage modulation)  were  chosen  in order to optimize the quality of the signal. In particular the large gate voltage modulations tend to washout  the main Dirac peak as well as dHvA oscillations.
 One can  identify the diamagnetic satellite peaks which are clearly split on the electron side  as well as the outer paramagnetic peaks. 
Similar data is also shown on $M_A$ in a smaller range of gate voltage  focusing on the regions in the vicinity of satellite Dirac points both in  hole and electron doping sides. In both cases paramagnetic peaks on both sides of the diamagnetic satellite peaks are visible with a more complex behavior with split peaks on the hole  side which will be discussed below.
\begin{figure*}[!ht]
\centering
\includegraphics[width=0.5\linewidth]{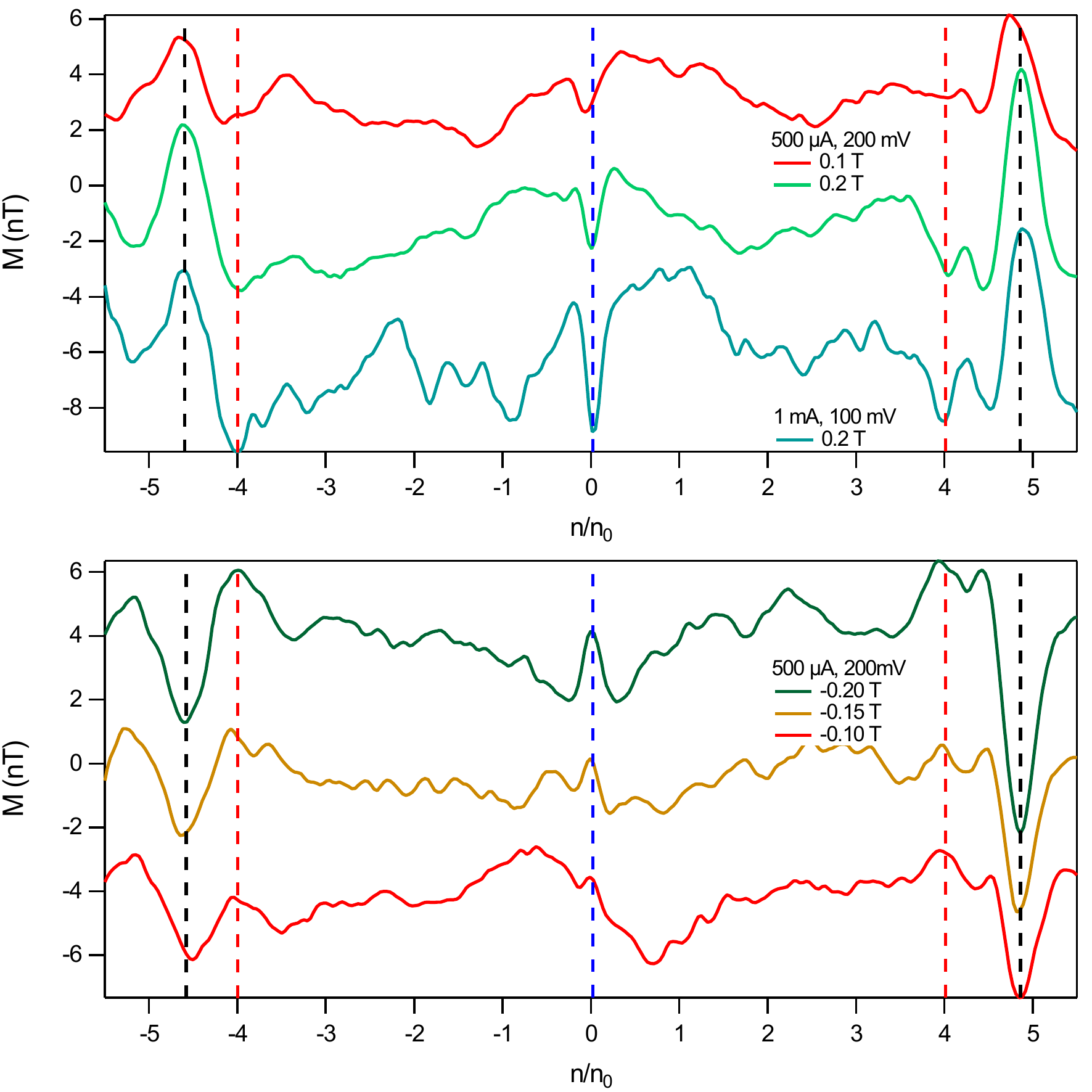}
\caption{Magnetization as a function of density (normalized by $n_0$) for sample $M_B$ for a set of additional low fields, different from the fields discussed in the main text. The top plot shows positive fields and the bottom plot shows negative fields. Curves have been shifted in the vertical axis for better visualization. For each curve the dc current though the GMR sensor is indicated as well as the amplitude of  modulation. Dashed vertical line indicate the paramagnetic Vignale peaks (black) and the diamagnetic McClure peaks.  }  
\label{PeaksMBFields}
\end{figure*}

\begin{figure*}[!ht]
\centering
\includegraphics[width=0.5\linewidth]{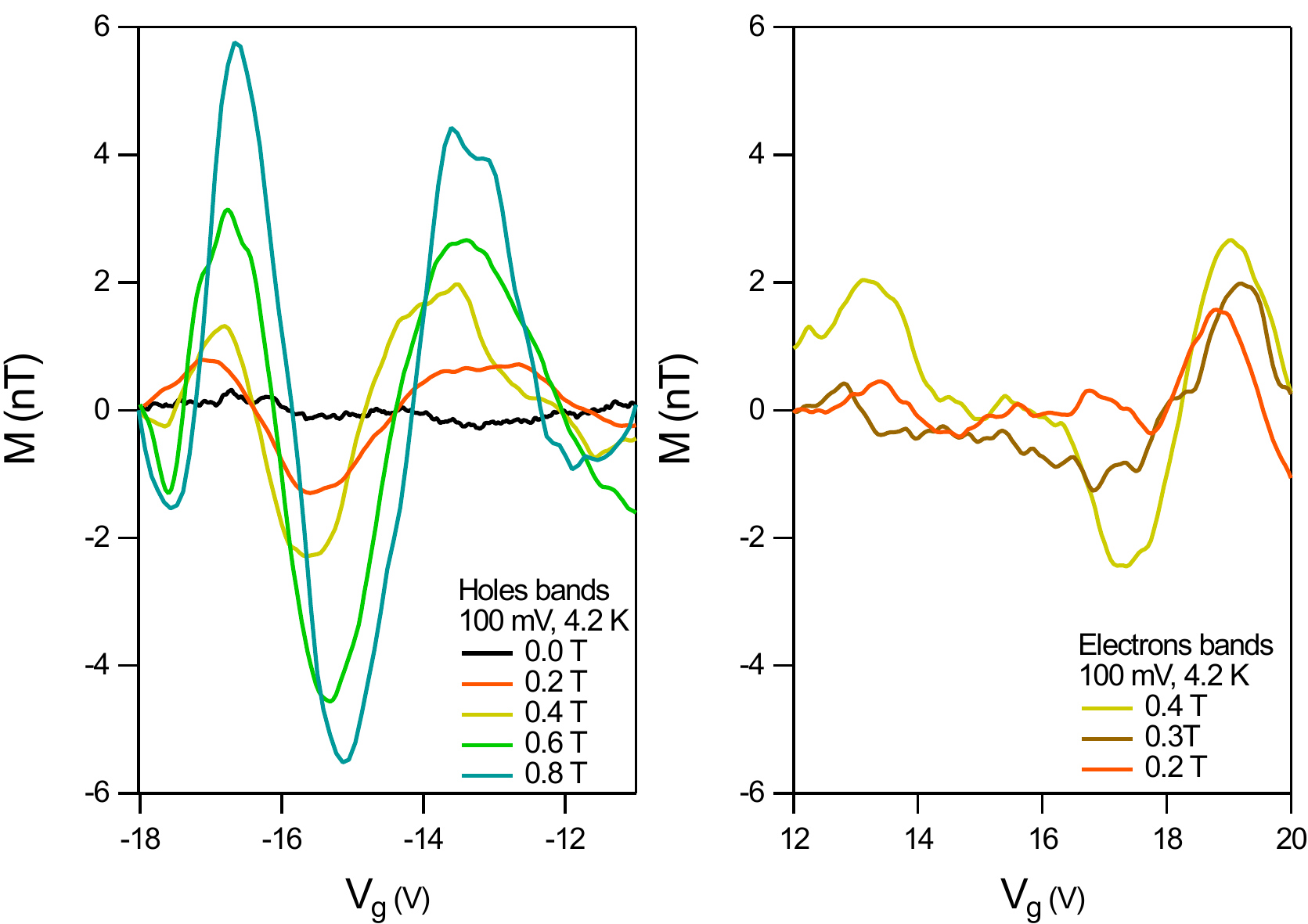}
\caption{Magnetization as a function of gate voltage for sample $M_A$ for different magnetic fields. The data shows only the vicinity of the secondary Dirac peaks in the hole and electron doping sides.}  
\label{FigSupMatMaHoles}
\end{figure*}

\begin{figure} 
\centering
    \includegraphics[width=0.5\linewidth]{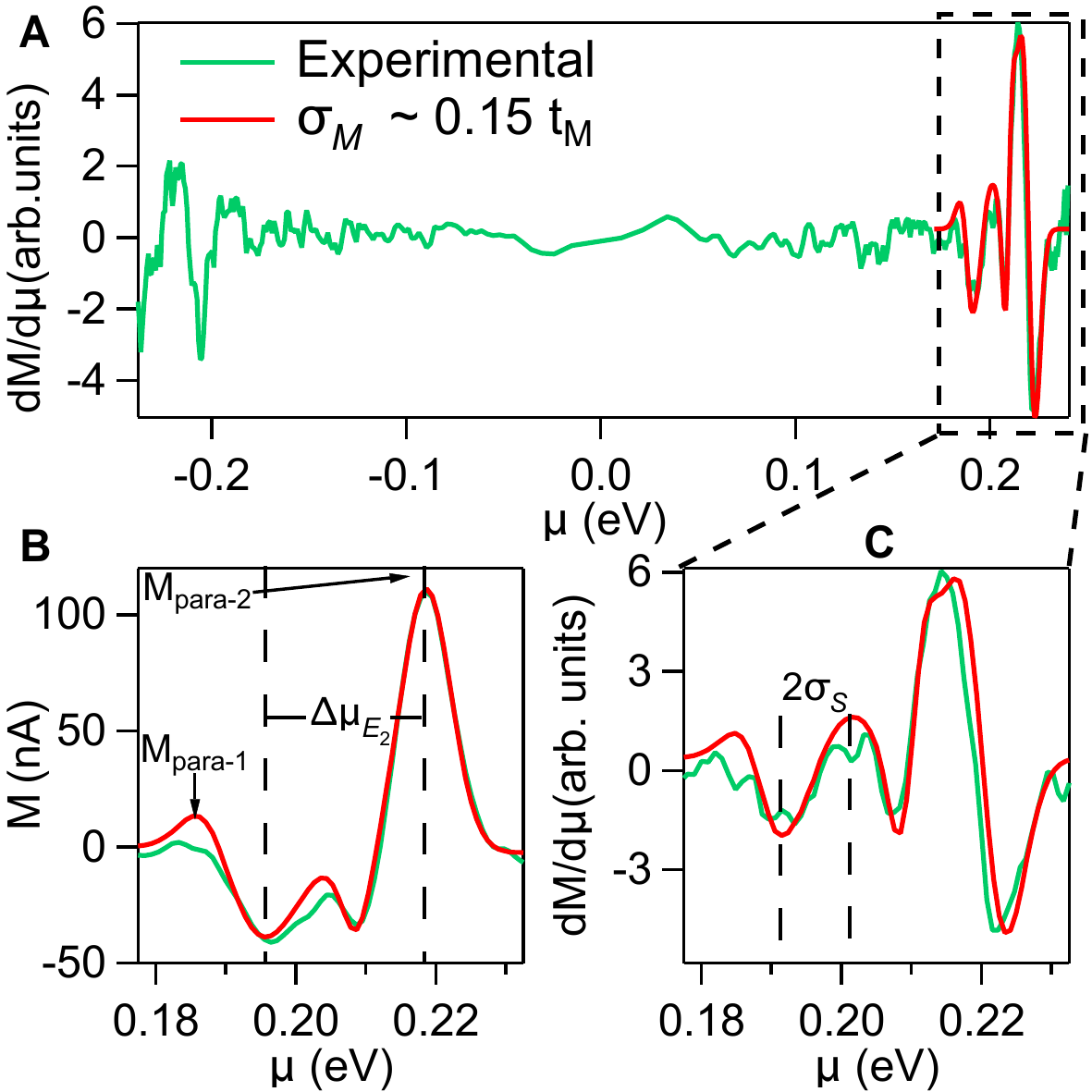}
    \caption{(A) In green, derivative of the magnetization as a function of the chemical potential for sample $M_B$. In red,  the signal is fit with the derivative of  4  gaussian peaks (2 paramagnetic and 2 diamagnetic) centered  respectively around the saddle points and the  split satellite Dirac peak. (B) Integral of the zone in a dashed rectangle in figure (A) as well as the integral of the fit. (C) Zoom of the rectangle zone in figure (A). From the fit of the experimental data we obtain  $\sigma_S=4.5 \pm 0.5 $ millielectronvolt.}
    \label{fig:bonus}
\end{figure}

%This might provide an  additional explanation  why it is more difficult in our experiment to observe the  paramagnetic peaks in vHs related to bands $E_1$ and $H_1$,  compared to  $E_2$ and $H_2$ for sample $M_B$. 
 Finally, in Fig.\ref{fig:bonus}, we explain how we determined $\sigma_S$, the amplitude of  disorder around  the satellite Dirac peaks  of the   hBN/Graphene bilayer of sample $M_B$. 
The value of $\sigma_S$,   is obtained from the distance in energy between maxima and minima of the chemical potential derivative of the secondary diamagnetic peaks as explained above. 

Similar analysis on sample $M_A$ gives a similar value of $\sigma_S=5.0 \pm 0.5 $ millielectronvolt.

\subsection{ Characteristics of the susceptibility peaks around the satellite Dirac points in comparison with numerical calculations}

In the following we present a more detailed analysis of the experimental data, than to what is done in the main paper,  in comparison with the band structure ans its analysis depicted in the previous section. For each sample investigated this analysis relies on only 2 adjustable parameters. The  first one is the energy width of the peaks estimated above and the second one is the amplitude of the moiré potential $t_M$ determined from the relative  positions of the satellite Dirac points and the positions of those large doping paramagnetic peaks clearly  which can be identified. 
\begin{figure*}[!ht]
\centering
\includegraphics[width=0.65\linewidth]{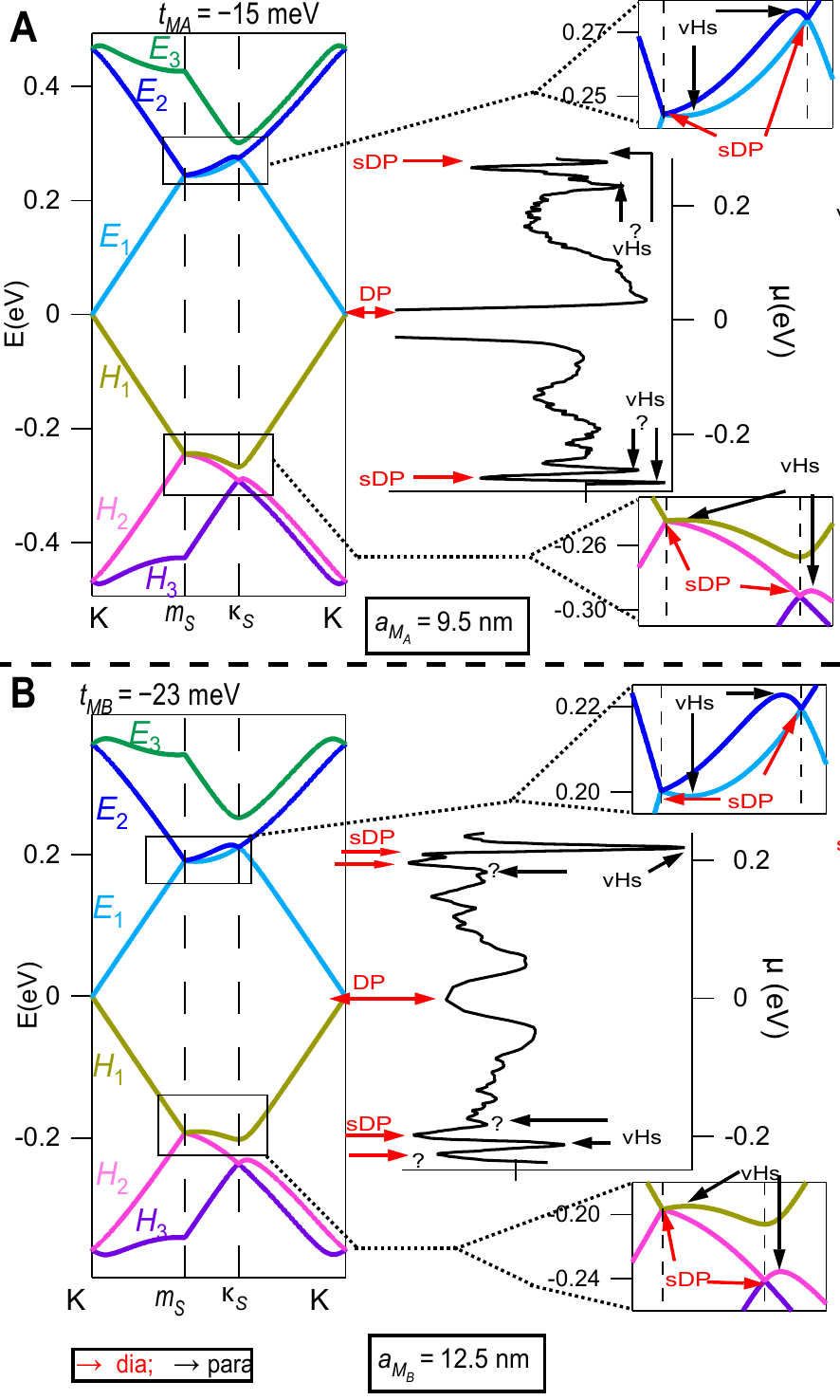}
\caption{ \textbf{(A)} and \textbf{(B)}  Magnetisation data on samples $M_B$ and $M_A$ measured at 0.2 T function of the  chemical potential in eV.  (We  note  that due to  the  $\sqrt{V_g}$ dependence, the energy width of satellite Dirac points is  smaller than the width of the main Dirac point.)
The difference between the chemical potential of  the sDP of the samples $M_B$ and $M_A$ is directly related to their different periodicity $a_M = 12.5 $ nm and $9.5$ nm respectively. Experimental data are compared to  cuts along the $K,\kappa_S$, $\kappa_S, m_S$, $m_S,K$ axis of the moiré band structure calculated for different values of $t_M$  matching the position of the observed diamagnetic peaks (red arrows) and paramagnetic peaks  (black arrows) of both samples. 
\label{4fig5_OMMSM}}
\end{figure*}
The next step is  then to understand the amplitudes of the different  peaks observed. In tables \ref{table:MaExp} and \ref{table:MbExp}, we summarize the parameters extracted and used for the comparisons between experimental and theoretical data.   Since our model does not take into account  the electron-hole asymmetry  in the experiment, one has to consider that the parameter $t_M$ is an average between the values which could be deduced fitting only the hole or the electron side of the data shown in Fig.4. $\Delta{\mu_{i}}_{E,H} \equiv\vert \mu_{para,i}-\mu_{dia,i}\vert_{E,H}$ (in  millielectronvolt ) are obtained from the distance in energy between the chemical potentials of the paramagnetic and corresponding diamagnetic peaks for each band (when they are visible), the rather large error bar comes from the width $\sigma_S$ which is not negligible compared to their spacing. The subindex $i=1,2$ is related to the considered  band : $E_{1,2}$ or $H_{1,2}$. The amplitudes of the magnetization $M_i$ (in  nanotesla) are measured directly from  the magnetization data. All values are obtained at an applied field of $0.2 $ tesla.
 From the  values  given in these tables, we can calculate the experimental value of $\frac{\chi_V}{|\chi_S|}$.  These ratios are of the same order of magnitude than the theoretical ratio $\frac{\chi_V}{|\chi_S|}$ for both samples discussed in the previous section.

\begin{table}[!ht]
    \centering
    \begin{tabular}{|c||c|c|c|c|c|c|c|c|c|}
    \hline
        $M_A$ & $\Delta\mu_{1}$ & $\Delta\mu_{2}$& $\sigma_S$ & $\vert M_{CNP}\vert$ & $\vert M_{sDP}\vert$ & $\vert M_{para-1}\vert$ & $\vert M_{para-2}\vert$\\ [0.5ex] \hline\hline
        $E_{1,2}$  & 10 $\pm $ 10 & 10 $\pm $ 2 & 4.8 & 15 & 1 $\pm $ 0.5 & 0.5 $\pm $ 0.5 & 1 $\pm $ 0.5 \\ \hline
        $H_{1,2}$ & 5 $\pm $ 5 & 12 $\pm $ 3 & 5 & " & 1.6 $\pm $ 0.5 & 1 $\pm $ 0.5  & 2 $\pm $ 0.5\\ \hline
    \end{tabular}
    \caption{Experimental parameters for sample $M_A$. $\Delta \mu_{i}=\vert \mu_{para,i}-\mu_{dia,i}\vert$ and $\sigma_S$ are given in  millielectronvolt . The amplitudes of the magnetization at charge neutrality point $M_{CNP}$, at the secondary Dirac points $M_{sDP}$, and the paramagnetic peaks $M_{para-1,2}$ are given in  nanotesla. The large errors bars come the small energy separation  between these peaks compared to their width. }
    \label{table:MaExp}
\end{table}

\begin{table}[!ht]
    \centering
    \begin{tabular}{|c||c|c|c|c|c|c|c|c|c|}
    \hline
        $M_B$ & $\Delta\mu_{1}$& $\Delta\mu_{2}$ & $\sigma_S$ & $\vert M_{CNP}\vert$ & $\vert M_{sDP}\vert$ & $\vert M_{para-1}\vert$  & $\vert M_{para-2}\vert$\\ [0.5ex] \hline\hline
       $E_{1,2}$  & N.A.& 23 & 4 & 5.6 & 1.8  & N.A. & 3.3 \\ \hline
       $H_{1,2}$  & N.A.& 14 & 5 & " & 1.7  & N.A. & 1.7\\ \hline
    \end{tabular}
    \caption{Experimental parameters for sample $M_B$. $\Delta\mu_{i}=\vert \mu_{para,i}-\mu_{dia,i}\vert$ and $\sigma_S$ are given in  millielectronvolt . The amplitudes of the magnetization at charge neutrality point $M_{CNP}$, at the secondary Dirac points $M_{sDP}$, and the paramagnetic peaks $M_{para-1,2}$ are given in  nanotesla.    \label{table:MbExp}}
\end{table}

In the following we show on two examples that it is possible to go further and reproduce the shape of the  magnetic singularities on the electron side considering  the contribution of {\it both diamagnetic peaks at Dirac points}   $ m_S$ and  $\kappa_S$ together with the paramagnetic sigularities at the saddle points which tend to compensate the diamagnetic ones. In particular we see that the proximity between the low doping saddle point  and $m_S$ has a different manifestation on both samples.
In the case of  $M_B$ the low doping(inner) paramagnetic Vignale peak  is largely reduced by the diamagnetic peak  of larger amplitude. 
This is why it is strongly depressed whereas the diamagnetic peak is only reduced by a factor two.
On the other hand, in the case of $M_A$ the inner paramagnetic peak  is still visible, due to its greater amplitude compared to the diamagnetic one at $m_S$,  which is in contrast strongly depressed whereas the diamagnetic peak at $K_S$ is clearly visible. 
We finally note that the outer paramagnetic peak is clearly visible on both samples.

\begin{figure*}[!ht]
\centering
\includegraphics[width=0.4\linewidth]{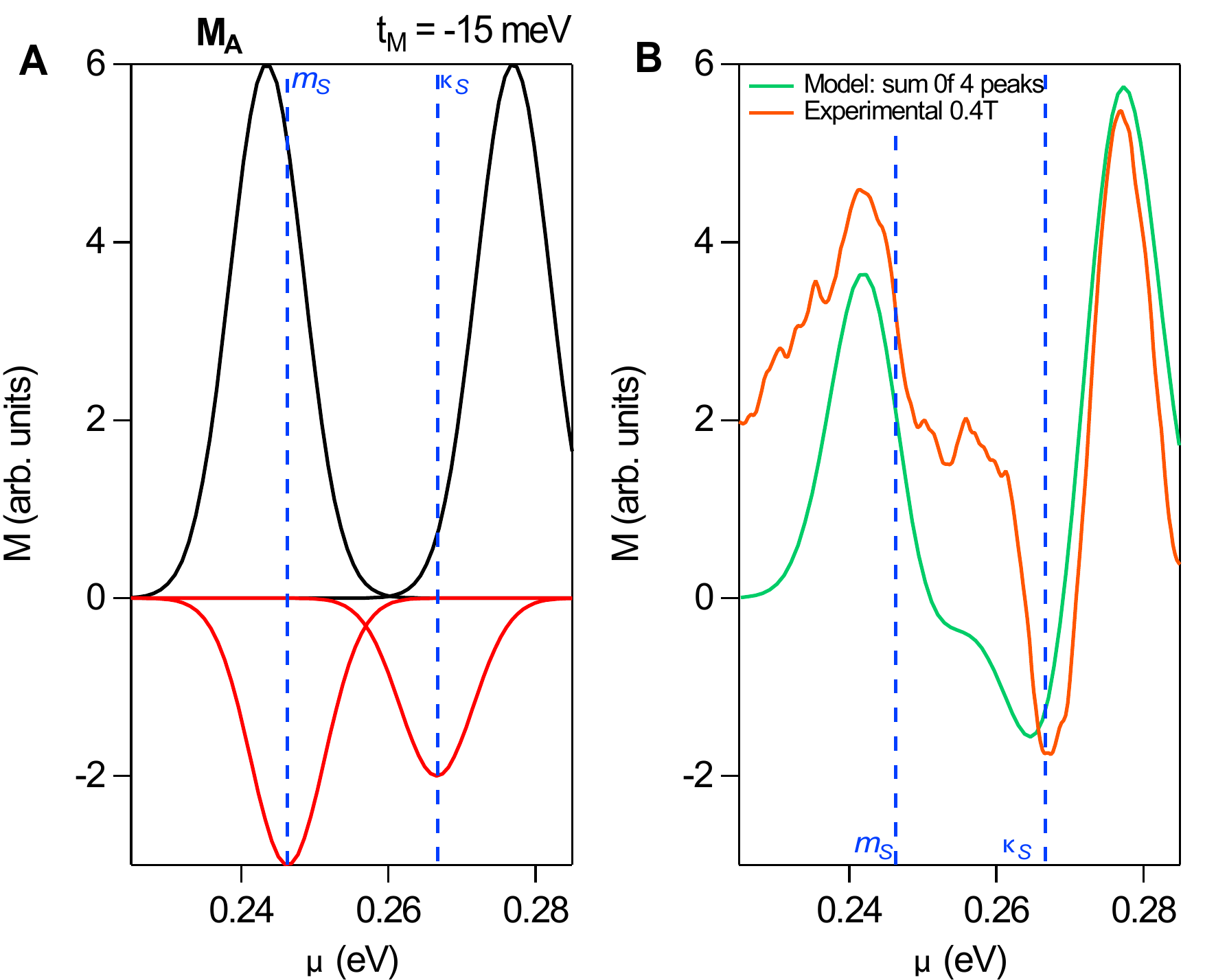}
\caption{\textbf{(A)} Illustration of how diamagnetic and paramagnetic responses combine on the  $H_1$ band for the $M_A$ sample.  We consider four similar separated peaks (two paramagnetic (black) and two diamagnetic (red)), centered at the energies determined by the band structure  shown in \ref{4fig5_OMMSM}A. These peaks have been constructed as gaussians with $\sigma_S=5$meV. Their amplitudes are determined by taking into account their multiplicity and the ratio $\chi_V/\vert\chi_S\vert=1 $ from the estimations in section I. \textbf{(B)} Comparison between the experimental data (orange) and the sum (green) of the 4 peaks in \textbf{(A)}. 
\label{SumPeaksMa}}
\end{figure*}

\begin{figure*}[!ht]
\centering
\includegraphics[width=0.4\linewidth]{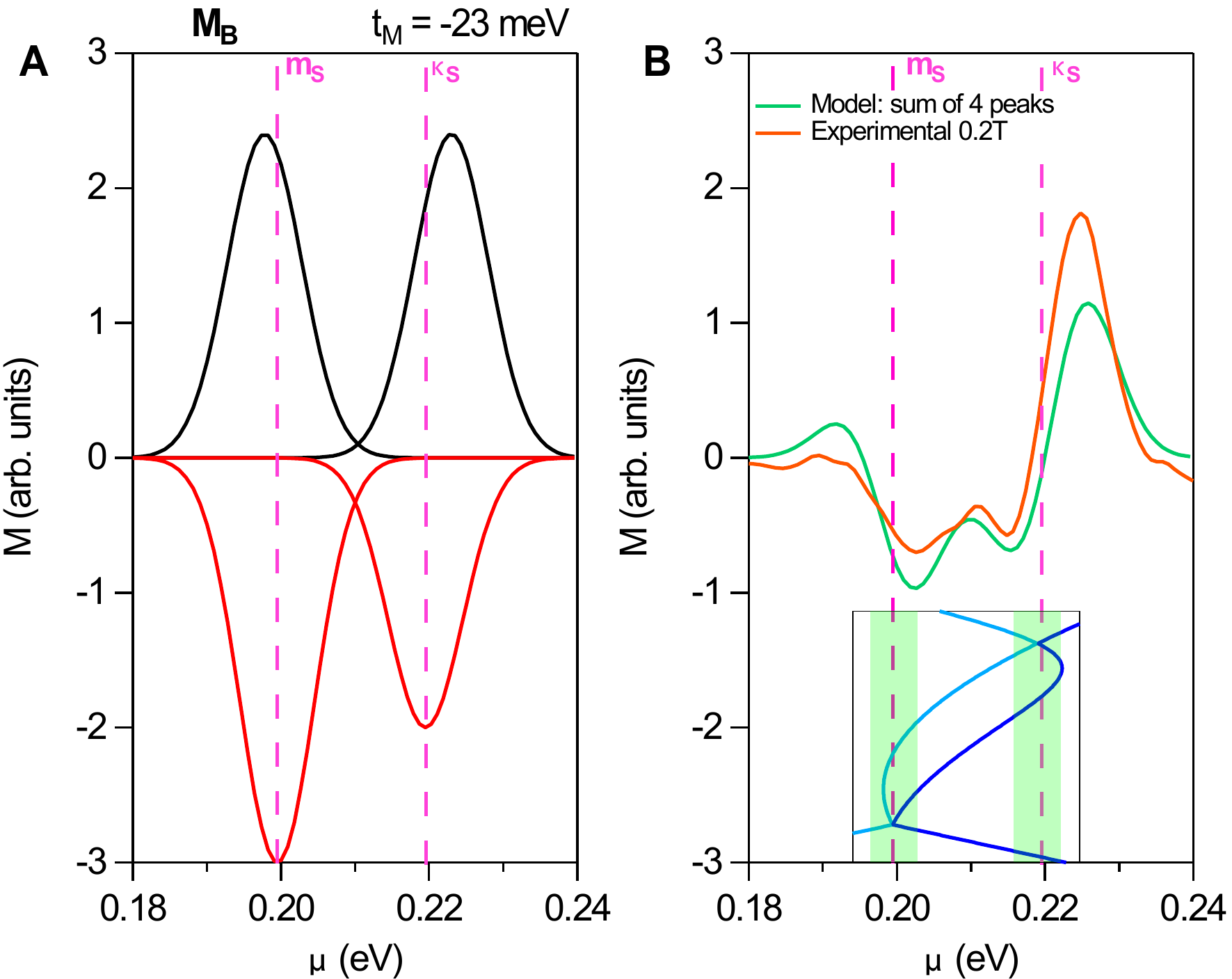}
\caption{\textbf{(A)}Illustration of how diamagnetic and paramagnetic responses combine on the  $H_1$ band for the $M_B$ sample.  We consider four similar separated peaks (two paramagnetic (black) and two diamagnetic (red))centered at the energies determined by the band structure in \ref{4fig5_OMMSM}B. These peaks have been constructed as gaussians with $\sigma_S=5$meV. Their amplitudes are determined by taking into account their multiplicity and the ratio $\chi_V/\vert\chi_S\vert=0.4$ from the etimation in section I. \textbf{(B)} Comparison between the experimental data (orange) and the sum (green) of the 4 peaks in \textbf{(A)}. The inset shows the related zone of the band structure, and  the green area shows the zone where the Dirac peak is broadened. 
\label{SumPeaksMB}}
\end{figure*}

%The situation is different on sample $M_A$ where the paramagnetic peaks are  clearly  visible on on both sides of the sDP, i.e.   on $H_1$ and $E_1$ bands as well as on $H_2$ and $E_2$ bands. On the other hand the  amplitude of the diamagnetic peak at the sDP  is lower than expected. This can be understood considering the small energy separation  between these peaks compared to their width and agrees with the parameters depicted in Table 1 pointing  towards a ratio $\chi_V/ \chi_M$ which is  larger for sample $M_A$ compared to $M_B$ by a factor 2.
%\subsubsection{Determination of $\sigma_S$}